\documentclass{aastex62}
\usepackage{lineno}

\usepackage[caption=false]{subfig}
\usepackage{graphicx}
\graphicspath{ {images/} }
\usepackage{hyperref}
\usepackage{amsmath}
\usepackage{booktabs}
\usepackage{xcolor}
\usepackage[percent]{overpic}
\usepackage{euler}
\newcommand\tantinu[1]{$\mathrm{\overline{\nu}_{#1}}$}
\newcommand\antinu{\overline{\nu}}
\newcommand\antinue{\overline{\nu}_e}
\newcommand\gammaray{$\mathrm{\gamma}$-ray}
\newcommand\tsigma{$\mathrm{\sigma}$}
\newcommand\Msun{$\mathrm{M_{\odot}}$}
\newcommand\alphaOri{$\mathrm{\alpha}$-Ori}
\newcommand\microsecond{$\mathrm{\mu s}$}

\begin{document}
\title{Sensitivity of Super-Kamiokande with Gadolinium to Low Energy Anti-neutrinos from Pre-supernova Emission}
\submitjournal{The Astrophysical Journal}
\received{2019-08-22}\revised{2019-09-22}\accepted{2019-09-25}

\newcommand{\AFFicrr}{\affiliation{Kamioka Observatory, Institute for Cosmic Ray Research, University of Tokyo, Kamioka, Gifu 506-1205, Japan}}
\newcommand{\AFFkashiwa}{\affiliation{Research Center for Cosmic Neutrinos, Institute for Cosmic Ray Research, University of Tokyo, Kashiwa, Chiba 277-8582, Japan}}
\newcommand{\AFFipmu}{\affiliation{Kavli Institute for the Physics and
Mathematics of the Universe (WPI), The University of Tokyo Institutes for Advanced Study,
University of Tokyo, Kashiwa, Chiba 277-8583, Japan }}
\newcommand{\AFFmad}{\affiliation{Department of Theoretical Physics, University Autonoma Madrid, 28049 Madrid, Spain}}
\newcommand{\AFFubc}{\affiliation{Department of Physics and Astronomy, University of British Columbia, Vancouver, BC, V6T1Z4, Canada}}
\newcommand{\AFFbu}{\affiliation{Department of Physics, Boston University, Boston, MA 02215, USA}}
\newcommand{\AFFuci}{\affiliation{Department of Physics and Astronomy, University of California, Irvine, Irvine, CA 92697-4575, USA }}
\newcommand{\AFFcsu}{\affiliation{Department of Physics, California State University, Dominguez Hills, Carson, CA 90747, USA}}
\newcommand{\AFFcnm}{\affiliation{Department of Physics, Chonnam National University, Kwangju 500-757, Korea}}
\newcommand{\AFFduke}{\affiliation{Department of Physics, Duke University, Durham NC 27708, USA}}
\newcommand{\AFFfukuoka}{\affiliation{Junior College, Fukuoka Institute of Technology, Fukuoka, Fukuoka 811-0295, Japan}}
\newcommand{\AFFgifu}{\affiliation{Department of Physics, Gifu University, Gifu, Gifu 501-1193, Japan}}
\newcommand{\AFFgist}{\affiliation{GIST College, Gwangju Institute of Science and Technology, Gwangju 500-712, Korea}}
\newcommand{\AFFuh}{\affiliation{Department of Physics and Astronomy, University of Hawaii, Honolulu, HI 96822, USA}}
\newcommand{\AFFicl}{\affiliation{Department of Physics, Imperial College London , London, SW7 2AZ, United Kingdom }}
\newcommand{\AFFkek}{\affiliation{High Energy Accelerator Research Organization (KEK), Tsukuba, Ibaraki 305-0801, Japan }}
\newcommand{\AFFkobe}{\affiliation{Department of Physics, Kobe University, Kobe, Hyogo 657-8501, Japan}}
\newcommand{\AFFkyoto}{\affiliation{Department of Physics, Kyoto University, Kyoto, Kyoto 606-8502, Japan}}
\newcommand{\AFFliv}{\affiliation{Department of Physics, University of Liverpool, Liverpool, L69 7ZE, United Kingdom}}
\newcommand{\AFFmiyagi}{\affiliation{Department of Physics, Miyagi University of Education, Sendai, Miyagi 980-0845, Japan}}
\newcommand{\AFFnagoya}{\affiliation{Institute for Space-Earth Environmental Research, Nagoya University, Nagoya, Aichi 464-8602, Japan}}
\newcommand{\AFFkmi}{\affiliation{Kobayashi-Maskawa Institute for the Origin of Particles and the Universe, Nagoya University, Nagoya, Aichi 464-8602, Japan}}
\newcommand{\AFFpol}{\affiliation{National Centre For Nuclear Research, 02-093 Warsaw, Poland}}
\newcommand{\AFFsuny}{\affiliation{Department of Physics and Astronomy, State University of New York at Stony Brook, NY 11794-3800, USA}}
\newcommand{\AFFokayama}{\affiliation{Department of Physics, Okayama University, Okayama, Okayama 700-8530, Japan }}
\newcommand{\AFFosaka}{\affiliation{Department of Physics, Osaka University, Toyonaka, Osaka 560-0043, Japan}}
\newcommand{\AFFox}{\affiliation{Department of Physics, Oxford University, Oxford, OX1 3PU, United Kingdom}}
\newcommand{\AFFqmul}{\affiliation{School of Physics and Astronomy, Queen Mary University of London, London, E1 4NS, United Kingdom}}
\newcommand{\AFFregina}{\affiliation{Department of Physics, University of Regina, 3737 Wascana Parkway, Regina, SK, S4SOA2, Canada}}
\newcommand{\AFFseoul}{\affiliation{Department of Physics, Seoul National University, Seoul 151-742, Korea}}
\newcommand{\AFFsheff}{\affiliation{Department of Physics and Astronomy, University of Sheffield, S3 7RH, Sheffield, United Kingdom}}
\newcommand{\AFFshizuokasc}{\affiliation{Department of Informatics in
Social Welfare, Shizuoka University of Welfare, Yaizu, Shizuoka, 425-8611, Japan}}
\newcommand{\AFFstfc}{\affiliation{STFC, Rutherford Appleton Laboratory, Harwell Oxford, and Daresbury Laboratory, Warrington, OX11 0QX, United Kingdom}}
\newcommand{\AFFskk}{\affiliation{Department of Physics, Sungkyunkwan University, Suwon 440-746, Korea}}
\newcommand{\AFFtokyo}{\affiliation{The University of Tokyo, Bunkyo, Tokyo 113-0033, Japan }}
\newcommand{\AFFtodai}{\affiliation{Department of Physics, University of Tokyo, Bunkyo, Tokyo 113-0033, Japan }}
\newcommand{\AFFtit}{\affiliation{Department of Physics,Tokyo Institute of Technology, Meguro, Tokyo 152-8551, Japan }}
\newcommand{\AFFtus}{\affiliation{Department of Physics, Faculty of Science and Technology, Tokyo University of Science, Noda, Chiba 278-8510, Japan }}
\newcommand{\AFFtoronto}{\affiliation{Department of Physics, University of Toronto, ON, M5S 1A7, Canada }}
\newcommand{\AFFtriumf}{\affiliation{TRIUMF, 4004 Wesbrook Mall, Vancouver, BC, V6T2A3, Canada }}
\newcommand{\AFFtokai}{\affiliation{Department of Physics, Tokai University, Hiratsuka, Kanagawa 259-1292, Japan}}
\newcommand{\AFFtsinghua}{\affiliation{Department of Engineering Physics, Tsinghua University, Beijing, 100084, China}}
\newcommand{\AFFynu}{\affiliation{Faculty of Engineering, Yokohama National University, Yokohama, Kanagawa, 240-8501, Japan}}
\newcommand{\AFFllr}{\affiliation{Ecole Polytechnique, IN2P3-CNRS, Laboratoire Leprince-Ringuet, F-91120 Palaiseau, France }}
\newcommand{\AFFbari}{\affiliation{ Dipartimento Interuniversitario di Fisica, INFN Sezione di Bari and Universit\`a e Politecnico di Bari, I-70125, Bari, Italy}}
\newcommand{\AFFnapoli}{\affiliation{Dipartimento di Fisica, INFN Sezione di Napoli and Universit\`a di Napoli, I-80126, Napoli, Italy}}
\newcommand{\AFFroma}{\affiliation{INFN Sezione di Roma and Universit\`a di Roma ``La Sapienza'', I-00185, Roma, Italy}}
\newcommand{\AFFpadova}{\affiliation{Dipartimento di Fisica, INFN Sezione di Padova and Universit\`a di Padova, I-35131, Padova, Italy}}
\newcommand{\AFFkeio}{\affiliation{Department of Physics, Keio University, Yokohama, Kanagawa, 223-8522, Japan}}
\newcommand{\AFFwinnipeg}{\affiliation{Department of Physics, University of Winnipeg, MB R3J 3L8, Canada }}
\newcommand{\AFFkcl}{\affiliation{Department of Physics, King's College London, London,WC2R 2LS, UK }}

\AFFox
\AFFipmu

\AFFicrr
\AFFkashiwa
\AFFmad
\AFFbu
\AFFuci
\AFFcsu
\AFFcnm
\AFFduke
\AFFllr
\AFFfukuoka
\AFFgifu
\AFFgist
\AFFuh
\AFFicl
\AFFbari
\AFFnapoli
\AFFpadova
\AFFroma
\AFFkcl
\AFFkeio
\AFFkek
\AFFkobe
\AFFkyoto
\AFFliv
\AFFmiyagi
\AFFnagoya
\AFFkmi
\AFFpol
\AFFsuny
\AFFokayama
\AFFosaka
\AFFregina
\AFFseoul
\AFFsheff
\AFFshizuokasc
\AFFstfc
\AFFskk
\AFFtokai
\AFFtokyo
\AFFtodai
\AFFtit
\AFFtus
\AFFtoronto
\AFFtriumf
\AFFtsinghua
\AFFwinnipeg
\AFFynu

\author{C.~Simpson}
\AFFox
\AFFipmu

\author{K.~Abe}
\AFFicrr
\AFFipmu
\author{C.~Bronner}
\AFFicrr
\author{Y.~Hayato}
\AFFicrr
\AFFipmu
\author{M.~Ikeda}
\AFFicrr
\author{H.~Ito}
\AFFicrr 
\author{K.~Iyogi}
\AFFicrr
\author{J.~Kameda}
\AFFicrr
\AFFipmu
\author{Y.~Kataoka}
\AFFicrr
\author{Y.~Kato}
\AFFicrr
% ...


\collaboration{The Super-Kamiokande Collaboration}
\noaffiliation

\shorttitle{Pre-SN \tantinu{e} at SK-Gd}
\shortauthors{The Super-Kamiokande Collaboration}

\begin{abstract}
Supernova detection is a major objective of the Super-Kamiokande (SK) experiment. 
In the next stage of SK (SK-Gd), gadolinium (Gd) sulfate will be added to the detector, which will improve the ability of the detector to identify neutrons. 
A core-collapse supernova will be preceded by an increasing flux of neutrinos and anti-neutrinos, from thermal and weak nuclear processes in the star, over a timescale of hours; some of which may be detected at SK-Gd.
This could provide an early warning of an imminent core-collapse supernova, hours earlier than the detection of the neutrinos from core collapse.
Electron anti-neutrino detection will rely on inverse beta decay events below the usual analysis energy threshold of SK, so Gd loading is vital to reduce backgrounds while maximising detection efficiency.
Assuming normal neutrino mass ordering, more than 200 events could be detected in the final 12~hours before core collapse for a 15-25 solar mass star at around 200~pc, which is representative of the nearest red supergiant to Earth, $\mathrm{\alpha}$-Ori (Betelgeuse).
At a statistical false alarm rate of 1~per~century, detection could be up to 10~hours before core collapse, and a pre-supernova star could be detected by SK-Gd up to 600~pc away.
\added{A pre-supernova alert could be provided to the astrophysics community following gadolinium loading.}\explain{Reviewer comment: + This paper focus on the sensitivity. If authors add the clear message about that SK-Gd will make alert system for the the astrophysical community, it is better.}

\end{abstract}
\correspondingauthor{C. Simpson}
\email{charles.simpson@physics.ox.ac.uk}

\section{Introduction}
	A core-collapse supernova (CCSN) produces a $\sim$10~second burst of neutrinos at tens of MeV, which is large enough that it can be detected by Super-Kamiokande (SK) and other neutrino detectors (\cite{Abe:2016waf}) if it is in or near the Milky Way.
	Much of what is known about galactic supernova explosions (SNe) comes from the detection of supernova (SN) neutrinos in 1987 (\cite{Hirata:1988ad,Bratton:1988ww,Alekseev:1988gp}).
	In the case of another SN in or near our galaxy, the current generation of neutrino detectors would be capable of collecting a much larger sample of SN neutrinos, improving our understanding and resolving outstanding questions about SNe.
	Neutrinos arrive before the electromagnetic radiation produced by a SN, so can generate a warning enabling astronomers to start observing as early as possible.
	Alert systems already exist for this purpose, linking together many detectors for maximum effect (\cite{Antonioli:2004zb}).
	SK has the currently unique ability of determining the direction of a SN from elastically scattered electrons, which is useful for guiding optical instruments (\cite{Abe:2016waf}); SN neutrino detection is a major goal for SK.

	Prior to collapse, as a star approaches the end of its life, the temperature and density increases, causing neutrinos from thermal and weak nuclear processes to become the main source of cooling as their emission is highly temperature dependent.
	For a nearby star, neutrino emission increases over a relatively short time-scale to detectable levels (\cite{OdrzywolekHeger}), which could give advanced warning before core collapse occurs.
	This earlier alert to the astronomy community could aid in observing the early light from a SN. 
	Advance warning could prevent SK (and other SN neutrino detectors) from missing a nearby core collapse due to planned down-time.
	SK has run for over 20~years so far, and has on average $>90\%$ up-time, making it especially useful for the detection of SNe, which are rare and could happen at any time. 
	\added{KamLAND, another neutrino detection experiment, has a functioning pre-SN alert system (\cite{Asakura:2015bga}).}\explain{Reviewer comment: Please add the KamLAND's preSN alarm system. Their system is already running. No mention is unnatural.}
	
	Furthermore, astrophysical neutrinos with known sources have only ever been detected from the Sun (for a review see \cite{Kirsten:1999bp}), from SN1987A, and blazar TXS 0506+056 (\cite{eaat1378}). 
	The only direct observation of SN neutrinos so far is SN1987A, and neutrino-cooled stars have never been observed, so detection would contribute to our understanding of late stage stellar evolution (\cite{Odrzywolek:2004conf,Kato:2015faa,Yoshida:2016imf,Patton:2017neq}).
	
	In the next stage of the Super-Kamiokande experiment (SK-Gd), gadolinium sulfate will be added to the detector, which will improve the ability of the detector to identify neutrons, and therefore low energy \tantinu{e} through inverse beta decay (IBD).
	Previous estimates \added{(\cite{Odrzywolek:2003vn,Patton:2017neq,Pablo:PhD})}\explain{Reviewer comment: Please add the reference about "previous estimates of SK's ability to detect pre-SN neutrinos"} of SK's ability to detect pre-SN neutrinos have assumed the energy threshold of the SK solar analysis (3.5~MeV positron kinetic energy) (\cite{Abe:2016nxk}). 
	In fact, detection will be possible at lower energies, albeit with reduced efficiency.
	
	In this article, the sensitivity of the SK-Gd detector to this pre-SN neutrino flux is assessed.
	The article is structured as follows. 
		Section \ref{Sec:PreSnNu} describes the pre-SN neutrino emission processes,
	Section \ref{Sec:SKGD} gives necessary information concerning SK-Gd,
	and theoretical pre-SN flux estimates are discussed in Section \ref{Sec:Flux}.
	The detection efficiency and backgrounds for low energy IBD are assessed in Section \ref{Sec:Detec}.
	Results of the study are presented in \ref{Sec:results}, followed by a conclusion in Section \ref{Sec:Conclusion}.

\section{Pre-supernova Neutrinos}\label{Sec:PreSnNu}
	In a massive star at the end of its life, the fusion of hydrogen~(H) and helium~(He) nuclei is insufficient to stabilise the temperature and density of the star. 
	That is, cooling from radiation is greater than heating from fusion, leading to contraction under gravity. 
	The higher density, and hence temperature, then enables the fusion of heavier nuclei, initially at the core of the star, and then in shells propagating outwards. 
	Neutrino emission is strongly temperature dependent.
	The increased temperature leads to an increased rate of cooling due to neutrino emission, which leads to further contraction and heating, and the fusion of even heavier nuclei (see \cite{Woosley:2002zz} for a review of late stellar evolution leading up to SNe). 
	This proceeds in stages driven by carbon~(C), neon~(Ne), oxygen~(O), and silicon~(Si) burning.
	Si burning creates an iron core, which can lead to a CCSN; these stars are sometimes called \emph{pre-supernova stars} (pre-SN). 
	Following the ignition of C burning, neutrino emission is the greatest source of cooling (\cite{Odrzywolek:2004conf}); so these stars are also referred to as \emph{neutrino-cooled stars}. 

	The neutrino-cooled stage of a massive star's life is remarkably short compared to usual astrophysical timescales, with the C burning stage lasting hundreds of years, Ne and O stages lasting under a year, and Si burning lasting under two weeks (\cite{Odrzywolek:2003vn}). 
	While these processes occur in the interior of the star, it may be that nothing changes in the star's electromagnetic emissions or at the outer surface of the star, so such a state would not be observed by electromagnetic astronomy (\cite{Odrzywolek:2004conf}). 

	Stars capable of CCSN are usually specified as having a zero age main sequence (ZAMS) mass $>$8~solar~masses~(\Msun{}), i.e. those capable of CCSN. 
	However, lighter stars in this category might not enter the Si burning stage, instead undergoing core collapse with an O-Ne-Mg core (\cite{Odrzywolek:2004conf}, \cite{Kato:2015faa}). 
		Very massive stars $>$30~\Msun{} may collapse directly to a black holes as \emph{failed supernovae}, but could still produce an increasing and detectable neutrino flux in late stages prior to collapse (\cite{Patton:2015sqt}). 
	Aside from ZAMS mass, stellar evolution also depends on metallicity and rotation; the models of pre-SN neutrino emission considered in this study assume solar metallicity and no rotation effects.

	Thermal processes, as well as $\beta^{-}$ decay and positron capture contribute to \tantinu{e} emission. 
	At these high temperatures a large equilibrium population of positrons exists, leading to neutrino emission by pair annihiliation $e^+e^-\rightarrow\antinu\nu$, which is important for detection prospects due to the high flux and relatively high average energy of the \tantinu{e} (\cite{Patton:2017neq,Odrzywolek:2004conf}). 

\begin{figure}
			\center
			\includegraphics[width=0.49\textwidth]{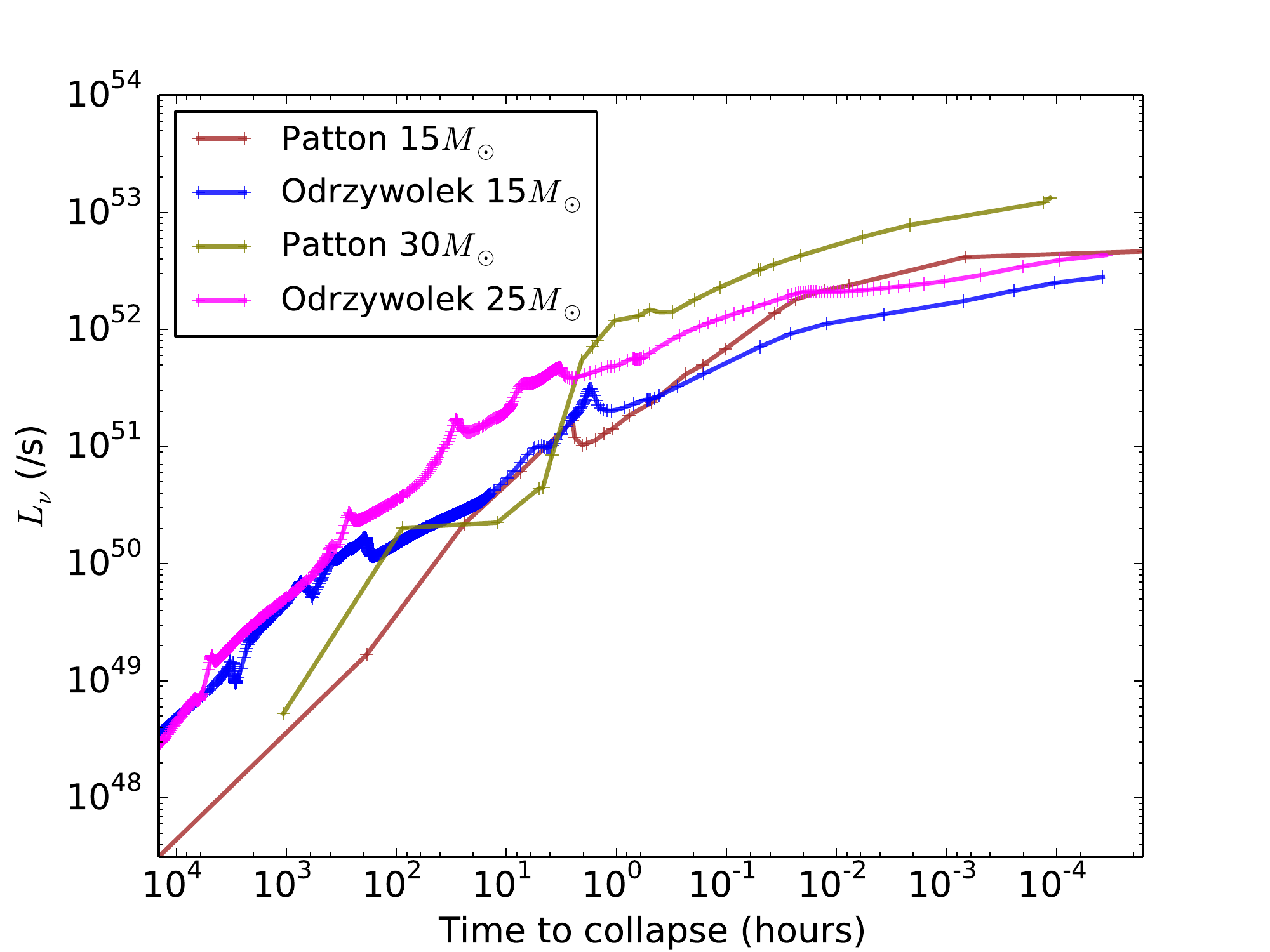}
			\includegraphics[width=0.49\textwidth]{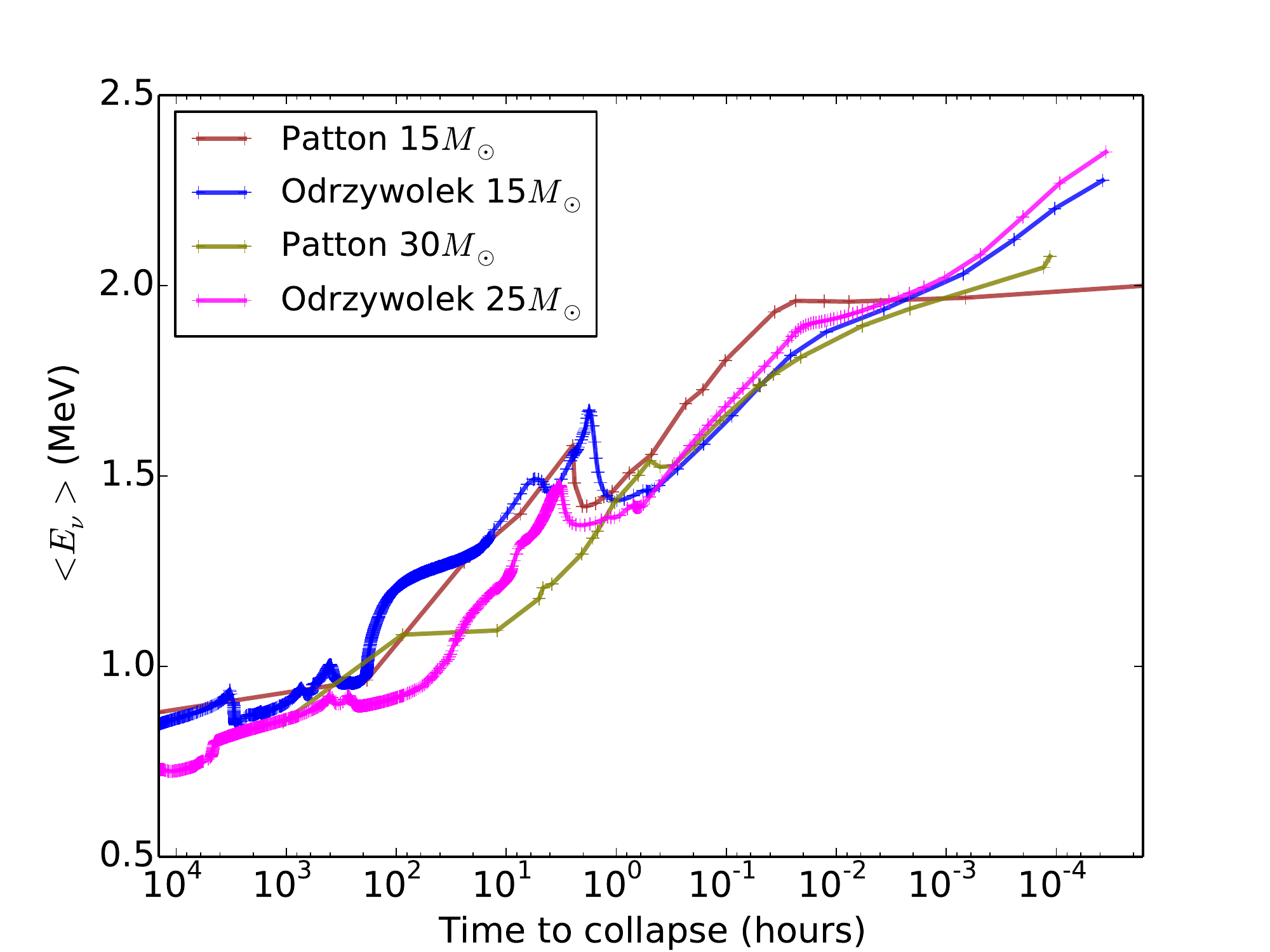}
			\caption{Comparison of $L_\nu$ number, and $<E_\nu>$ mean energy, of emitted \tantinu{e}, against time to core collapse, in models by \cite{OdrzywolekHeger} and \cite{Patton:2017neq}. 
			Spikes in both quantities are caused by ignition of fusion of heavier isotopes in the core or shell of the star.}
			\label{Fig:LE_nu_odr_pat}
	\end{figure}

\begin{figure}[htbp]
		\center
		\includegraphics[width=0.49\textwidth]{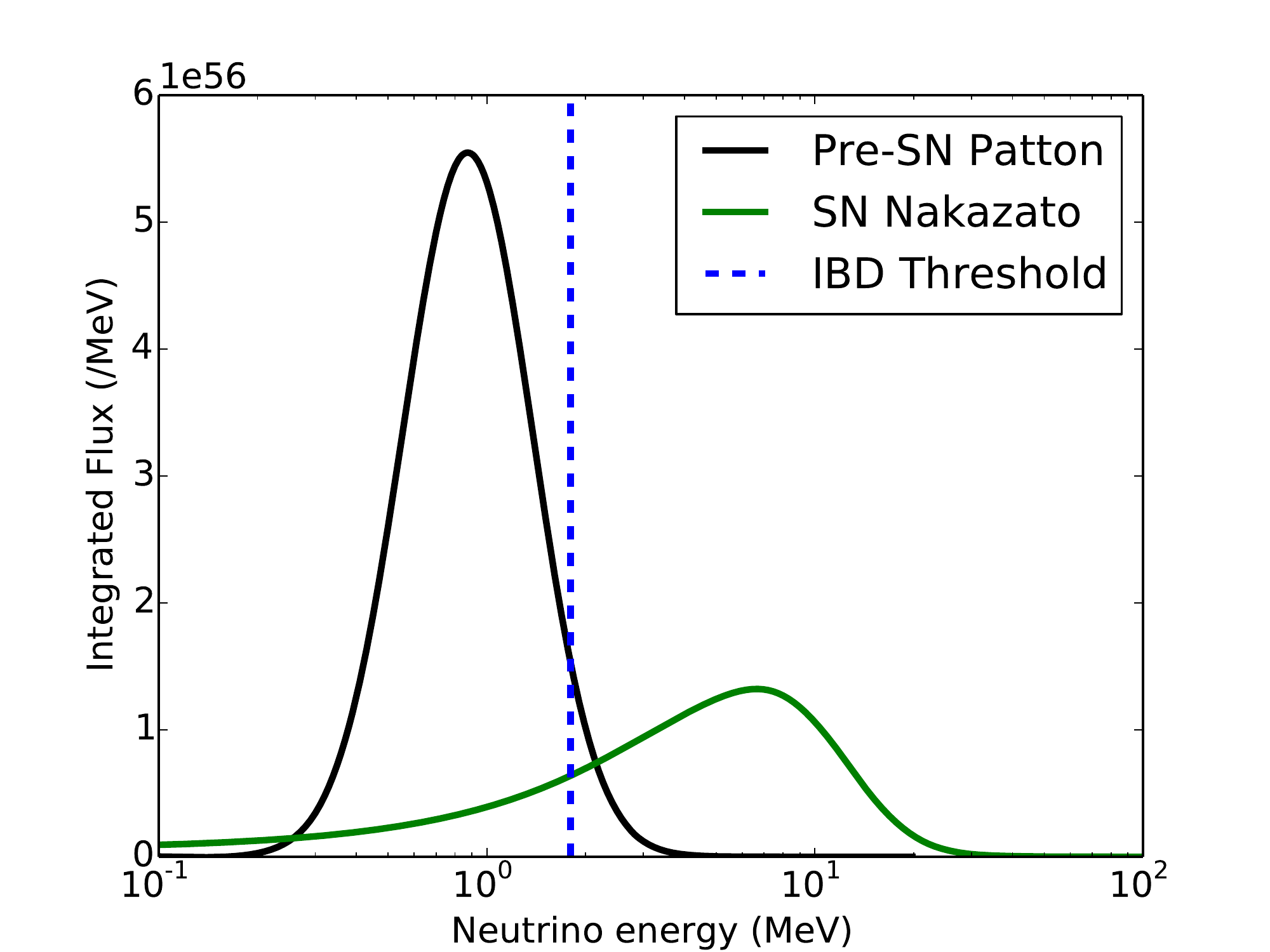}
		\caption{Total emission versus energy for \tantinu{e} from pre-SN and from a SN burst, integrated over the full time range of the respective models. 
		The dotted line shows reaction threshold of IBD. 
		Pre-SN spectrum shown is a 30~\Msun{} model from \cite{Patton:2017neq}, 
		SN spectrum is 30~\Msun{} from \cite{Nakazato:2012qf}. 
		30~\Msun{} was chosen so that the two would have the same mass.
		Note that the pre-SN neutrinos are emitted over a much longer timescale (1000~hours vs. 20~seconds), and that the IBD cross section is strongly energy dependent.}
				\label{Fig:SNvsPreSN}
	\end{figure}

	SK-Gd has a chance of detecting a pre-SN star following the ignition of Si burning, as the rate of \tantinu{e} emission (\autoref{Fig:LE_nu_odr_pat}(a)), and crucially the \tantinu{e} average energy (\autoref{Fig:LE_nu_odr_pat}(b)), increase as the star approaches core collapse (\cite{OdrzywolekHeger}).
	As shown by \autoref{Fig:SNvsPreSN}, pre-SN \tantinu{e} emission is at much lower energies than those of a SN.
	Furthermore, pre-SN emission is over a very long timescale compared to SN emission. 
	The mean \tantinu{e} energy is below the IBD threshold, meaning only the tail could possibly be detected through IBD; the increasing average energy will mean that the proportion above threshold will also increase.
	For a sufficiently nearby star, SK-Gd would see a rapid increase in the rate of low energy IBD candidate events. This would presage a CCSN by hours. 

	From \cite{Nakamura:2016kkl}[Table 2], there are 41 red supergiant (RSG) stars with distance estimates within 1~kpc, 16 within 0.5~kpc, and 5 within 0.2~kpc (\cite{Nakamura:2016kkl}[Table 2]). 
	Famous nearby RSGs include Betelgeuse \alphaOri{}{}, Antares ($\mathrm{\alpha}$ Sco), and $\mathrm{\epsilon}$ Peg.
	Wolf-Rayet stars are also possible supernova progenitors, e.g. \replaced{$\mathrm{\gamma^2}$ Vel}{$\mathrm{\gamma}$ Vel~A}\explain{This is a more widely used notation for binaries.}.
	
\section{Super-Kamiokande with Gadolinium}\label{Sec:SKGD}
	SK is a large water Cherenkov detector, and is well described elsewhere (\cite{FUKUDA2003418,ABE2014253}). It consists mainly of a tank of 50~kilotons~(kt) of ultra-pure water. 
	The inner detector (ID) is 32~kt, and the fiducial volume (FV) is usually given as 22.5~kt, although in practice a smaller or larger FV is used by different analyses as appropriate. 
	In this paper quoted efficiencies assume the full ID volume.
	The ID is instrumented with around 11,000 50~cm photomultiplier tubes (PMTs). 
	Charged particles are detected through their emission of Cherenkov radiation.
	
	\emph{Super-Kamiokande with Gadolinium} (\emph{SK-Gd}, formerly \emph{GADZOOKS!}) is the next phase of the SK experiment. 
	The main aim of SK-Gd is to detect the supernova relic neutrino signal within a few years of adding gadolinium (Gd) (\cite{Beacom:2003nk}). 
	
	SK-Gd will add gadolinium sulfate ($\mathrm{Gd_2(SO_4)_3}$) to SK's pure water. 
	Naturally abundant isotopes of Gd have some of the highest thermal neutron capture (TNC) cross sections.
	TNC on Gd is followed by a \gammaray{} cascade with a total energy of $\sim$8~MeV, much more than the single 2.2~MeV \gammaray{} produced by TNC on hydrogen which is currently used at SK for neutron tagging (\cite{ZHANG201541}). 
	Mainly through Compton scattered electrons, \gammaray{}s~can be detected in SK indirectly.
	The \gammaray{} cascade from TNC on Gd produces visible energy comparable to an electron with $\sim$4~to~5~MeV total energy.

	The main channel for detection of \tantinu{e} at low energy (roughly $<10$~MeV) in SK is IBD on hydrogen ($\mathrm{H(\antinue,e^+)n}$), as its cross section is relatively high. 
	The neutron takes a short time to thermalise in water and capture, and travels only a short distance, meaning that the positron and TNC form a delayed coincidence (DC), in which two events are reconstructed within a short time and distance of each other. 
	This method of detection is made possible by the upgrade to QBee electronics described in \cite{DAQUpgrade,NISHINO2009710}.
	The probability of uncorrelated events producing this signature is low, so neutron tagging allows electron anti-neutrino events to be distinguished from background events, including neutrino events. 
	The high TNC cross section of Gd makes the time between the prompt and delayed parts of the event shorter than with H ($\sim$20~\microsecond{} vs. $\sim$180\microsecond{}), 
		and the higher visible energy improves the vertex reconstruction resolution. 
	As a result, tagging efficiency for signal will be higher, and accidental backgrounds lower. 

	Note that in low energy IBD, the direction of the incoming \tantinu{e} cannot be reconstructed from the direction of the emitted positron (\cite{Vogel:1999zy}), and the number of elastic scattering events will be small for a pre-SN, so this technique will not have any SN pointing ability.
	\added{The direction of neutron travel cannot be resolved in SK, so positron-neutron vector cannot be used to infer the \tantinu{e} direction either.}\explain{Reviewer comment:Note for directional detection (page 5) is true. However, it might make missunderstand. ﻿Delayed-prompt vector is sensitive to the direction of incoming neutrinos.}

	It is planned that SK's ultra-pure water will be loaded with gadolinium sulfate in two steps, firstly to 0.02\% by mass, then to 0.2\%; leading to 50\% and 90\% of neutrons capturing on Gd respectively, with the rest mainly capturing on H.
	This paper assumes 0.2\% Gd loading, so it should be noted that SK-Gd will begin with a period of reduced sensitivity. 
	
	Research and development of the required technologies for SK-Gd has been undertaken by the EGADS experiment, which has successfully operated a Gd-loaded water Cherenkov detector for over two years at 0.2\% gadolinium sulfate loading (\cite{EGADS}). 	
	
\section{Electron Anti-neutrino Flux}\label{Sec:Flux}
\begin{figure}
		\center
		\gridline{
			\fig{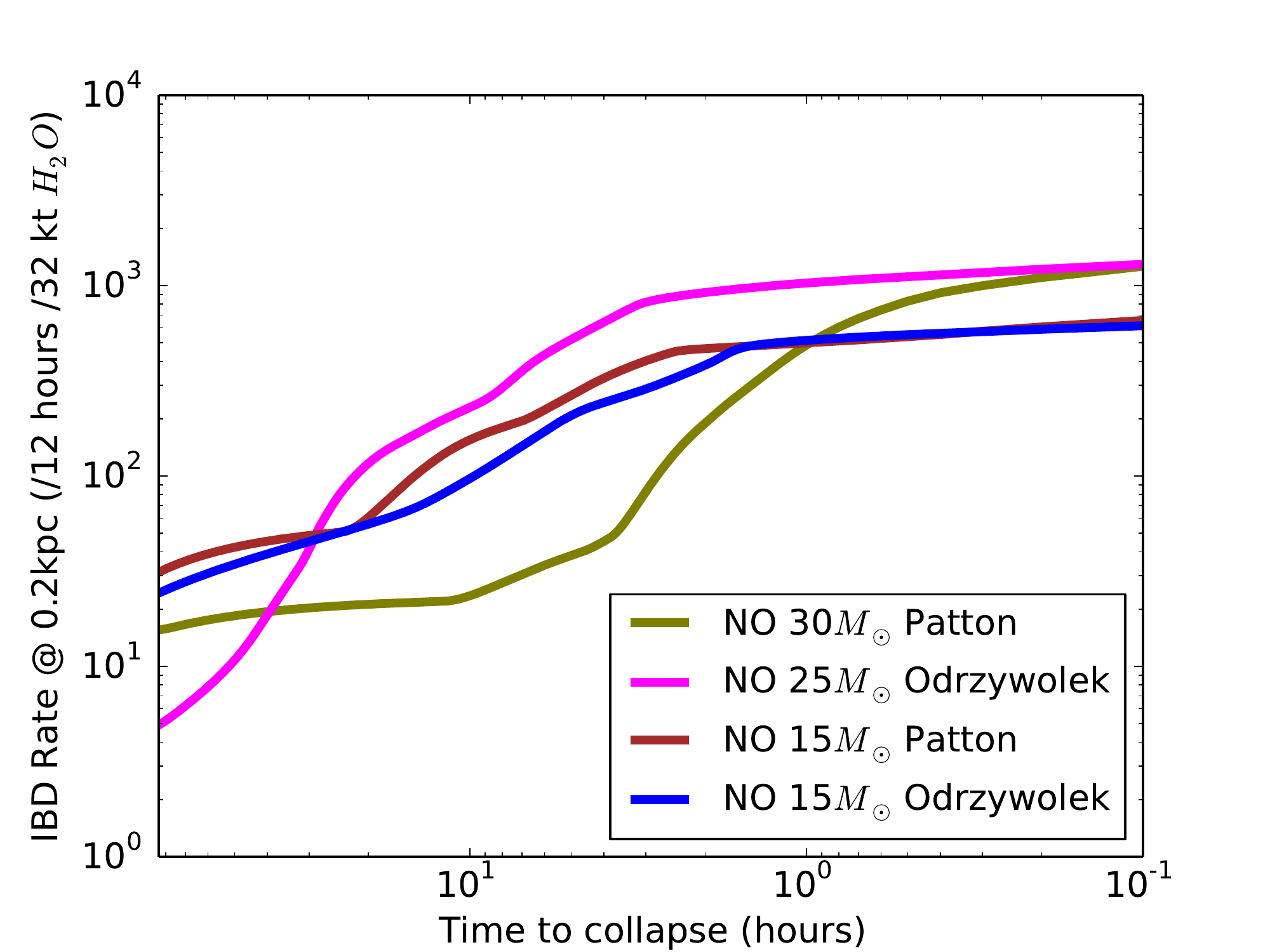}{0.49\textwidth}{(a) IBD rate per 32~kton vs. time in a 12~hour integration window}
			\fig{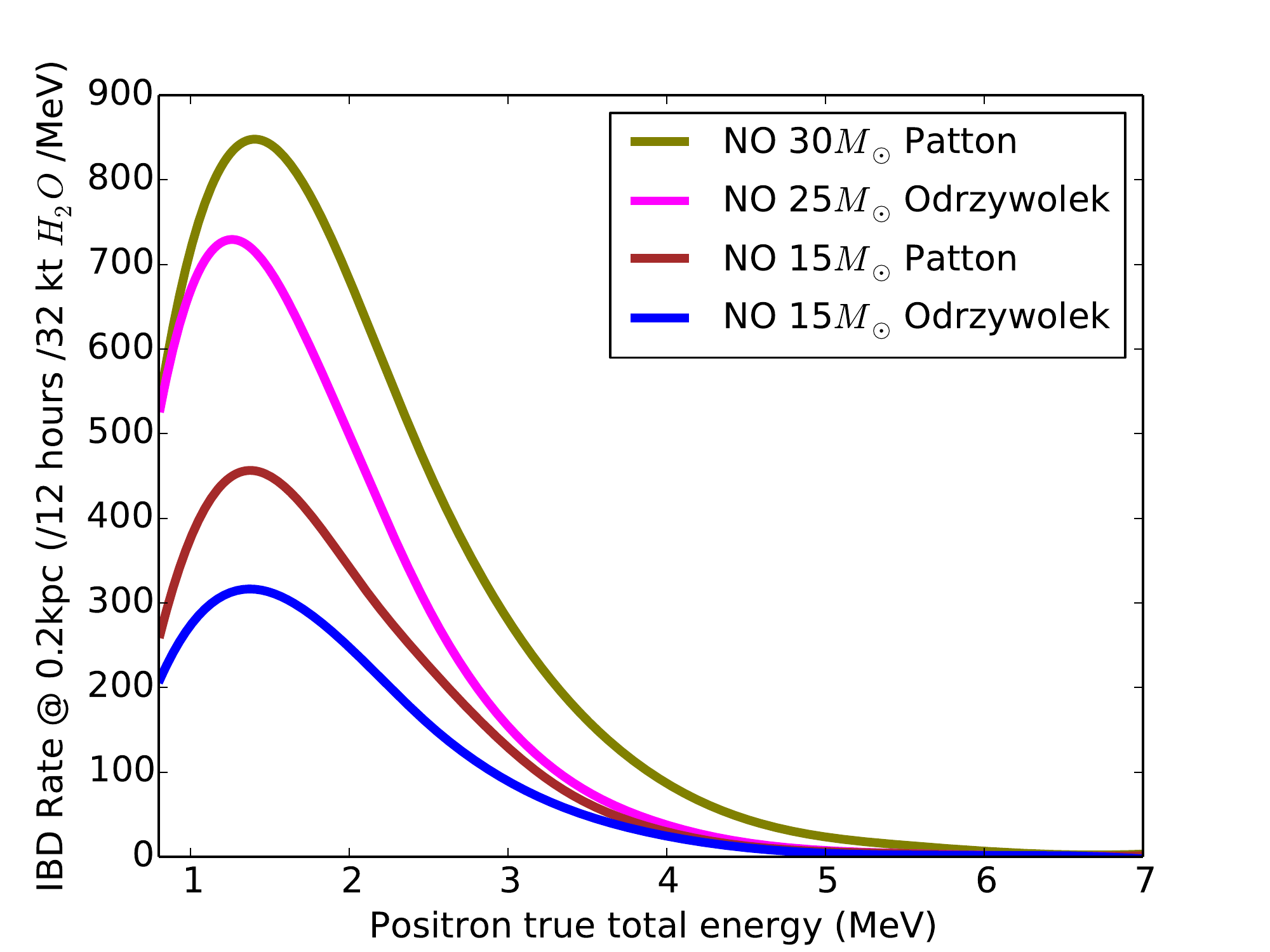}{0.49\textwidth}{(b) IBDs in 32~kton in final 24~hours before CCSN. positron true total energy, only shown above Cherenkov threshold.}
		}
		\caption{
		Comparison of total number of IBDs in a 12~hour window per kt $\mathrm{H_2O}$ against time to core collapse, predicted in models by \cite{OdrzywolekHeger} and \cite{Patton:2017neq}. 
		Detection efficiency is not taken into account.
		A distance of 200~pc is assumed.
		Note that a reduced time range is used compared to \autoref{Fig:LE_nu_odr_pat}.
		Interpolation has been used to produce equally spaced time and energy points.}
		\label{Fig:flux_nu_odr_pat}
	\end{figure}

	Neutrino emissions are calculated from stellar models.
	Although there are several sets of predictions published for the flux of \tantinu{e} from a pre-SN,
	this study primarily uses the datasets of \cite{OdrzywolekHeger} (data downloaded from \cite{Web:PSNS}) and \cite{Patton:2017neq} (data downloaded from \cite{patton_kelly_m_2019_2626645}).
	\cite{Patton:2017neq} predict similar \tantinu{e} total emission rates to \cite{OdrzywolekHeger} for a 15\Msun{} star, as shown in \autoref{Fig:LE_nu_odr_pat} and \autoref{Fig:flux_nu_odr_pat}, though the time and energy dependent emission rates differ.
	The flux estimates of \cite{OdrzywolekHeger} are calculated by post-processing the output of an already existing stellar model, and isotopic composition is calculated by assuming nuclear statistical equilibrium (\cite{2009PhRvC..80d5801O}).
	\cite{Patton:2017neq,Patton:2015sqt} use a more modern stellar evolution code, which fully couples the isotopic composition to the stellar evolution, and tracks the rates of a larger number of isotopes individually.
	Isotopic composition especially affects the rate of neutrinos produced by weak nuclear processes.

	Time $t=0$ is taken to be the moment at which the stellar simulation is terminated (when the infall velocity exceeds some threshold), which can be taken as the beginning of core collapse. 
	Figure \ref{Fig:flux_nu_odr_pat} contrasts the time dependent predicted IBD rates and the positron true energy spectra from the models considered.
		
	The electron flavour ratio of the neutrinos is affected as it passes through the dense stellar medium. 
	Following \cite{Patton:2017neq} and \cite{Kato:2017ehj}, an adiabatic transition is assumed, and the flavour is changed by the Mikheyev-Smirnov-Wolfenstein high resonance, dependent on the neutrino mass ordering (MO).
	The assumed transition probability is $P(\antinue \rightarrow \antinue) = 0.675(0.024)$ in the normal (inverted) mass ordering case, and \replaced{$P(\antinu_{x} \rightarrow \antinue) = 1 - P(\antinue \rightarrow \antinu_{x})$.}{$P(\antinu_{x} \rightarrow \antinue) = 1 - P(\antinue \rightarrow \antinu_{e})$.}\explain{Typo}
	In the data of \cite{Patton:2017neq}, the \tantinu{x} flux is provided. 
	For the data of \cite{OdrzywolekHeger}, the initial $\antinu_{x} / \antinue$ ratio assumed to be 0.19 following \cite{Asakura:2015bga}.
	This is assumed to not be energy or time dependent, and includes the effect of vaccum oscillations.
	Earth matter effects have not been included.
	Only the electron anti-neutrino flavour will interact through IBD, the rest of the flux is assumed to be invisible in this energy range at SK.

	From the flux as a function of neutrino true energy and time, the expected rate of IBD reactions at SK-Gd are calculated.
	\replaced{The effect of distance is simply a $1/R^2$ effect.}{The effect of distance is simply a factor $1/R^2$.}\explain{Repetition of effect.} An accurate approximation of the energy dependent cross section is used \cite[Eqn. 25]{Strumia:2003zx}. 
	The number of targets for IBD is the number of hydrogen nuclei in the SK ID. 
	Detection efficiency is dependent on the energy of the positron, and energy resolution is taken into account with a smearing matrix calculated from detector simulation.

\section{Detection Efficiency and Backgrounds} \label{Sec:Detec}
	\subsection{Energy Thresholds at SK-Gd}\label{Sec:Thresh}
		The most fundamental energy thresholds for SK-Gd are $E_\nu >$1.8~MeV for IBD, and $E_e >$ 0.8~MeV total energy for Cherenkov emission by positrons. 
		To a good approximation, the total energy $E_e$ of a positron produced by IBD is related to the electron anti-neutrino energy $E_\nu$ by $E_e=E_\nu-\Delta$, where $\Delta=m_n-m_p=1.293$~MeV is the nucleon mass difference.
		Positrons above the Cherenkov threshold may not be reconstructed if there are not enough photons detected. 
		Cherenkov photons detection inefficiencies arise due to attenuation in the water, the photocathode coverage of the detector, and the quantum and collection efficiencies of the PMTs (\cite{FUKUDA2003418}).

		\begin{figure}[htbp]
			\center
						\includegraphics[width=0.49\textwidth]{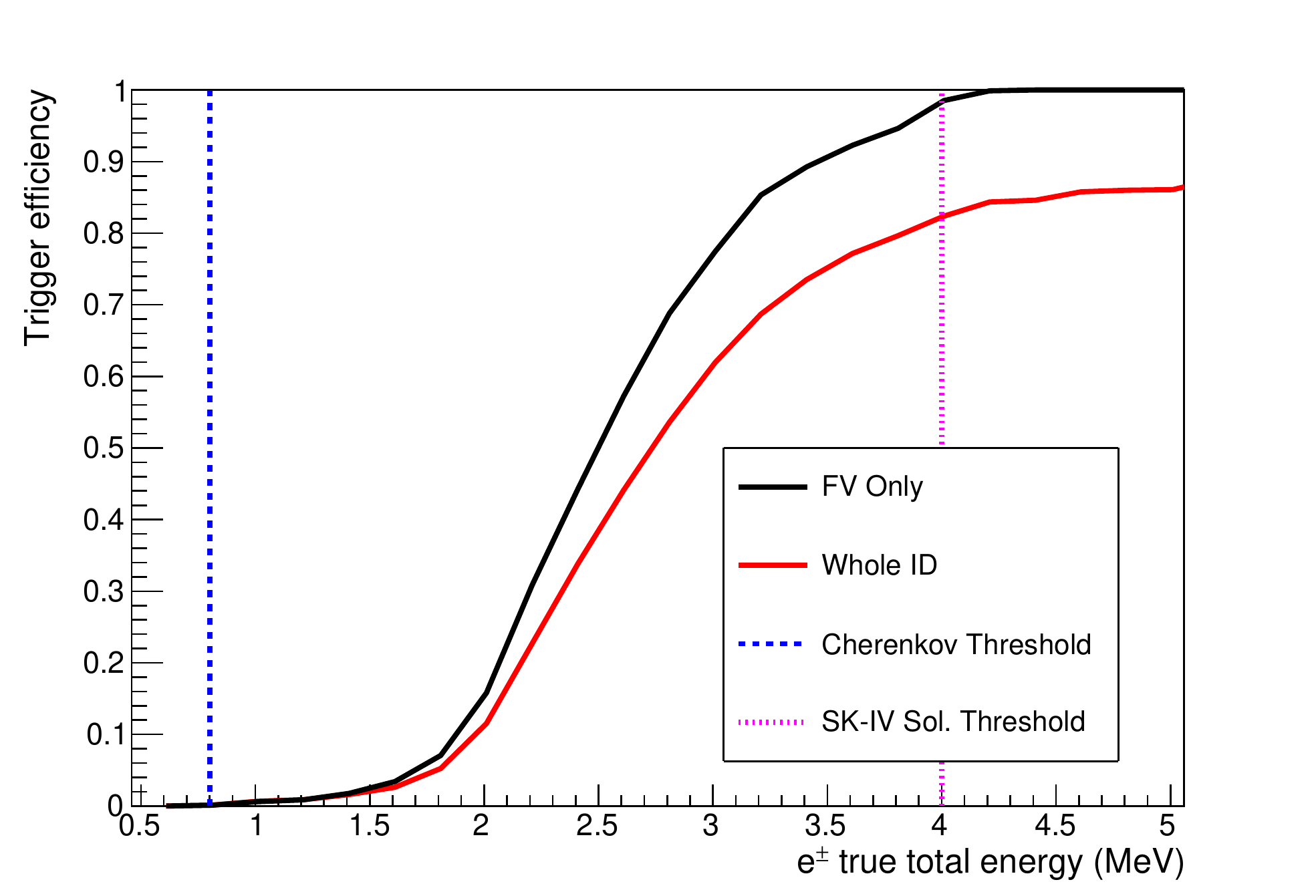}
			\caption{WIT efficiency against electron true total energy, evaluated with simulation. 
			Note that true energy is shown, not reconstructed energy.
			5000 MC events were generated for each 0.2~MeV true energy bin.	\added{The efficiency for events generated in the whole ID is compared to those generated only inside the FV.}\explain{Reviewer comment: + Figure 4. Is "FV Only" the  inside of fiducial volume or only fiducial volume cut ? Please make it clear.}
			The Cherenkov threshold and SK-IV solar neutrino analysis threshold (4~MeV total electron energy \cite{Abe:2016nxk}) are shown for reference. 
			}
			\label{Fig:eEff}
		\end{figure}
		In SK, a PMT is considered \emph{hit} if the charge collected by that PMT exceeds a threshold.
				In order to reduce the data rate from dark noise and radioactivity, a threshold is usually applied to the number of hits in a time window (\cite{Abe:2016nxk}).
		The \emph{Wide-band Intelligent Trigger} (WIT) is an independent trigger system at SK that uses parallel computing to reconstruct very low energy events which do not meet the usual thresholds. 
		It is close to 100\% efficient for electrons around 4~MeV total energy within the FV (\cite{Carminati:2015jna}).
		Below this, the efficiency drops, as shown by Figure \ref{Fig:eEff}. 

		There is no way to reliably distinguish positrons at very low energy ($<4$~MeV total) from the intrinsic radioactive background of SK, which increases by an order of magnitude for each MeV the energy threshold is lowered.
		Requiring DC with a TNC on Gd can make background rates to IBD manageable, even at the lowest energies SK can reconstruct.
		Furthermore, even if the IBD positron is not detected and successfully reconstructed, the \gammaray{} cascade from TNC on Gd can be detected. 
		Such \emph{single neutron events} will be subject to higher backgrounds than those in DC, but the threshold in neutrino energy is effectively reduced to the threshold for the IBD reaction ($E_\nu > 1.8 $~MeV). 

	\subsection{Signal Event Model}
		Low energy positrons in 0.2~MeV true total energy bins from 0.8 to 7~MeV, as well as \gammaray{} cascades resulting from TNC on Gd, were simulated using the standard detector simulation Monte Carlo of SK (\cite{Koshio:PhD}). 
	
		The spectrum of \gammaray{}s~from TNC on Gd consists of a few known energy \gammaray{}s~close to the Q-value, and a continuum at lower energies where the levels are so densely populated that they are indistinguishable. 
		How energy should be divided up between the \gammaray{}s~in each event is not well known, so approximate models are used.
		The SK detector simulation is based on GEANT3, which does not simulate TNC on Gd by default. 
		GEANT4 (\cite{Geant:AGOSTINELLI2003250, Geant:ALLISON2016186}) contains a number of generic models, but they do not contain those known high energy \gammaray{}s.
		For the GLG4SIM generic liquid scintillator simulator, another model was developed (\cite{Web:GLG4sim}), which combined a parametric model of the continuum with the known high energy \gammaray{}s. 
		Recent efforts at J-PARC (\cite{Das:2017jdy,Hagiwara:2017hxs,Ou:2014nva}) have used new measurements to account for what combinations of \gammaray{}s~come together, and have provided another model including this as well as a different continuum distribution. 
		The two latter models give similar distributions of the reconstructed variables, so it is likely that the detector is fairly insensitive to the details. 
		The model used by GLG4SIM was used as the basis for selection training and efficiency calculations, with that provided by the J-PARC group used as a cross check.
		The difference between the two models is included as an uncertainty.

		Real data from the SK detector were used as background noise, into which were injected the PMT hits produced by the simulated events.
		The hybrid data were then subjected to the same triggering algorithms used by the WIT system. 
		It was assumed that all events had to be independently reconstructed and selected by WIT. 
		\replaced{In the SK Collaboration ongoing efforts are being made to loosen the WIT requirements on reconstruction quality in the case that an event candidate is in DC with another, so in future trigger efficiencies for DC events at low energies could be significantly higher than those used in this study.}{In future, requirements on reconstruction quality could be loosened in the WIT for events in coincidence, or prompt and delayed vertices could be fit simultaneously, slightly improving triggering efficiencies for low energy prompt events.}\explain{Although our analyses are always being improved upon, the results presented in this article are correct and complete. 
			Work on the improved coincidence trigger is still preliminary, and a long way from being ready for publication.
			The sensitivity given in this article could be achieved by SK with only the addition of gadolinium sulfate, with no changes to our data acquisition, and minimal changes to our data processing streams.
			Furthermore, improvements to the efficiency will not be dramatic, affecting the overall sensitivity less than the uncertainty assumed on the background rate in this study.}

\begin{figure}[htbp]
		\centering
		\gridline{
									\begin{overpic}[trim={0.4cm 0.7cm 1.45cm 0},clip,width=0.5\textwidth]{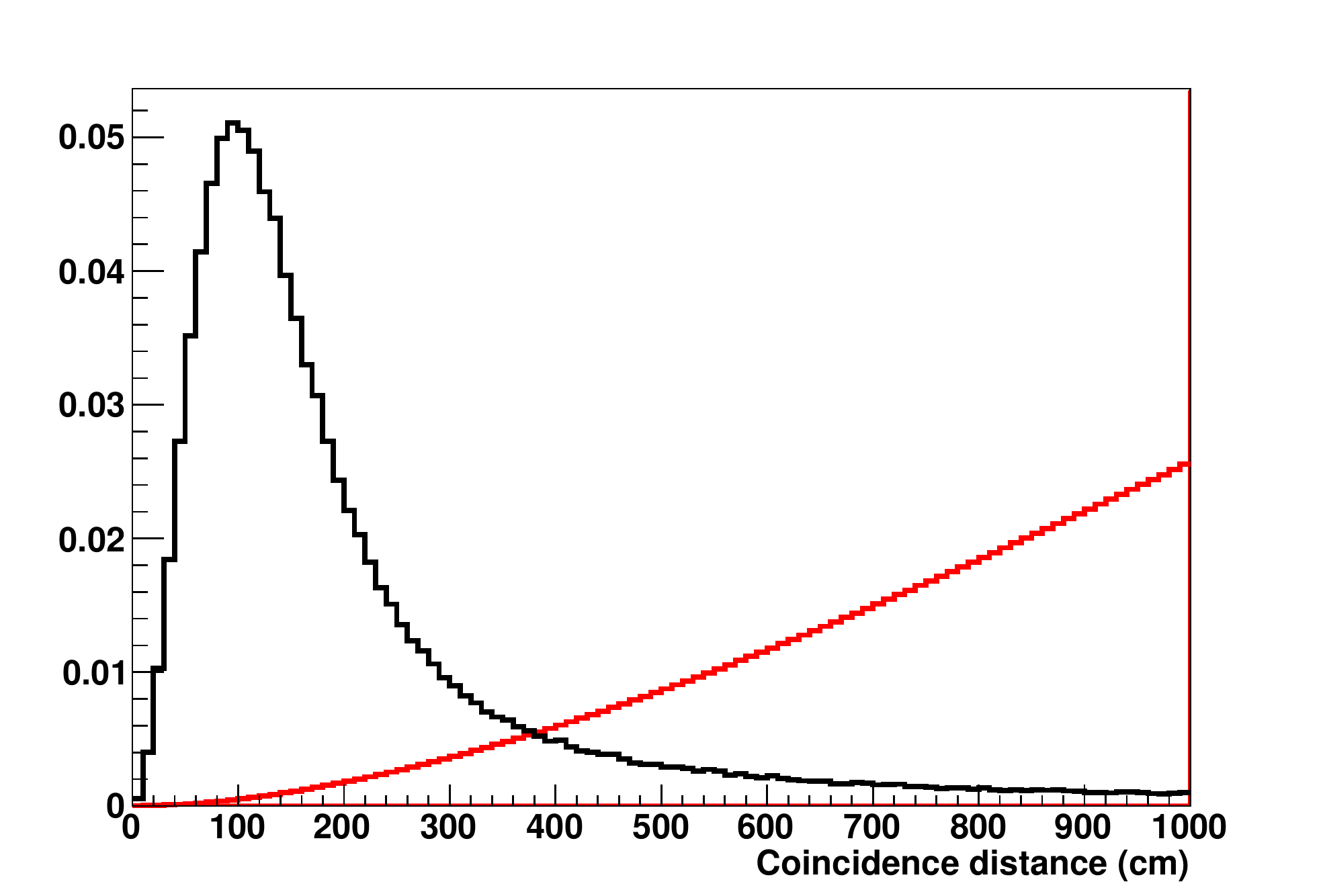}
								\put(50,-2){\makebox(0,0){\begin{tabular}{c}(a) \textbf{Coincidence distance} (cm)\end{tabular}}}
			\end{overpic}
			\begin{overpic}[trim={0 0.7cm 1.85cm 0},clip,width=0.5\textwidth]{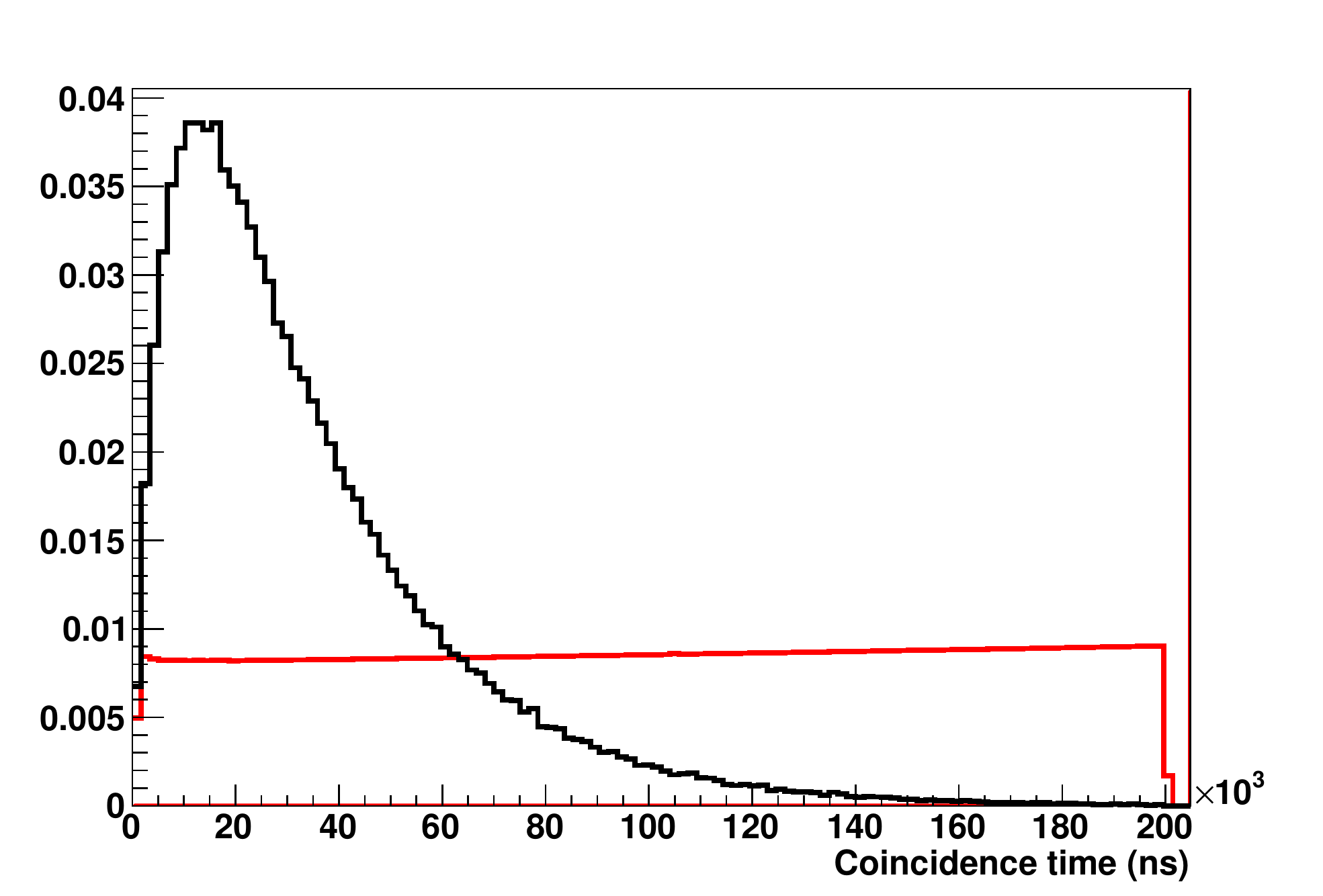}
				\put(50,-2){\makebox(0,0){\begin{tabular}{c}(b) \textbf{Coincidence time }($\mathrm{\mu s}$)\end{tabular}}}
			\end{overpic}
		}

		\caption{
			Distribution of the reconstructed distance (a) and reconstructed time (b) between simulated event candidates (with 3~MeV prompt positrons) before selection. 
			Signal MC is shown in black and accidental background from data in red, area normalised to one.
		}
		\label{Fig:coiTimDist}		
	\end{figure}

		The important features of a DC are the temporal and spatial separation of its two parts, shown in \autoref{Fig:coiTimDist}.
		The distribution of spatial separation is dependent on the energy of the positron through the vertex reconstruction resolution, and is modelled by pairing positrons and neutrons created at the same true locations in the detector simulation. 
		Neutron transport was not modelled, i.e. the neutrons were assumed to capture at the same point at which they were produced. 
		This is a reasonable assumption as, at these low energies, the distance travelled by neutrons before capturing (order 10~cm) is much smaller than the position reconstruction resolution (order 1~m). 
		The distribution of DC times was assumed to have the form $e^{-\frac{t}{\tau_c}}(1-e^{-\frac{t}{\tau_t}})$, where $\tau_t$ is the time taken for the neutron to thermalise, and $\tau_c$ the time for it to capture. 
		Thermalisation and capture times were set to values measured using an americium-beryllium neutron source in EGADS loaded to 0.2\% gadolinium sulfate by mass.
					
	\subsection{Background Model}

		The background to the single neutron channel is mostly events from dark noise and radioactive decays with similar characteristics to TNCs (``fake neutrons''), and to a lesser extent events which are real TNCs not from pre-SN neutrinos. 
		Backgrounds to DC type events include accidental DC between unrelated background events, real DC for certain radioactive decays, and real DC from reactor neutrinos and geoneutrinos. 
		
		It is anticipated that rates for intrinsic backgrounds will be known quite precisely once in-situ measurements become possible, but estimates have been made during the planning and development of SK-Gd.
		In order to estimate the rate of fake neutrons intrinsic to SK, events recorded by the WIT system during 6000 hours of normal pure water data from the fourth run period of SK (SK-IV) were used. 
		An initial event selection for neutrons was designed to reduce the rate of fake neutrons to a reasonable level, while retaining as much efficiency for simulated TNC events as possible. 
		These data were also used to estimate the accidental DC rate by searching for pairs of events in DC.
		
		A large part of the intrinsic background of the SK detector comes from radon emanation into the water from materials inside the detector and radioactive decays in the detector materials; this is mainly concentrated at the edges and bottom of the detector (\cite{TAKEUCHI1999418,Nakano:Radon}), and is reduced by fiducial volume and energy threshold cuts. 
		
		Any radioactive impurities left in the gadolinium sulfate will be distributed throughout the detector volume by loading. 
		Much of the work preparing for SK-Gd has been in quantifying these, including with low background counting using high-purity germanium detectors, and working with chemical manufacturers to reduce the level of contamination such that it does not detrimentally impact other SK analyses (\cite{EGADS}). 
		In this study, to allow for this additional contamination, backgrounds calculated from SK-IV data, and assumed backgrounds from $\mathrm{(\alpha,n)}$ and spontaneous fission (SF), are scaled up by a factor of two in the worst case.

		The process $\mathrm{^{18}O(\alpha,n)^{21}Ne^{*}}$, and its equivalent with $\mathrm{^{17}O}$, produce neutrons. 
		There was some concern that this could produce a high rate of neutrons, especially if there were a high contamination of $\mathrm{^{235}U}$ series isotopes that are $\alpha$ emitters. 
		Efforts to reduce radioisotope contamination of gadolinium sulfate have brought this down to an acceptable predicted level.
		As $\mathrm{^{21}Ne^{*}}$ decays by neutron emission, each reaction of this kind produces two neutrons, so it would be possible for one of the TNCs to be mistaken for a positron, creating a DC background.
		Rates were calculated using the SOURCES code \cite{Wilson:SOURCES4C}.
		At expected levels of contamination from the $\mathrm{^{238}U}$ ($<5~mBq/kg$), $\mathrm{^{235}U}$ ($<3~mBq/kg$), and $\mathrm{^{232}Th}$ ($<0.05~mBq/kg$) chains (\cite{EGADS}), we estimate these processes will contribute 6-12 pairs of neutrons per day before detection efficiency. 		
		SF of $\mathrm{^{238}U}$ also produces on average more than one neutron per fission (\cite{UCRL-AR-228518}), and \gammaray{}s~that could be mistaken for positrons, producing a DC background. 
		Assuming a $\mathrm{^{238}U}$ contamination of the gadolinium sulfate of 5~mBq/kg, the SF rate is calculated to be 0.6 per hour in the ID before efficiencies. 		The \gammaray{}s~are assumed to have falling energy distribution (\cite{PhysRevC.7.1564}).
		Using the neutron multiplicity from \cite{doi:10.13182/NSE07-85}, it is assumed that decays with more than one neutron can also form a DC.
		The contribution to the background of SF turns out to be subdominant in this analysis.
		Beta delayed neutrons from SF will be negligible.
				
		Reactor and geo electron anti-neutrinos are an irreducible background, being the same particles in the same energy range as the signal. 
		The reactor background depends strongly on the number of Japanese nuclear reactors that are running. 
		To account for this, the reactor and geo neutrino flux was calculated using the \url{geoneutrinos.org} web app (\cite{Barna:2015rza}), which combines reference models for reactor neutrinos (\cite{reactor_neutrinos_reference_model}) and geoneutrinos (\cite{geoneutrinos_reference_model}). 
		Reactor activity is assumed as the mean values given by IAEA's PRIS database for the years 2010 and 2017 (\cite{PRIS}).
		During 2010 most Japanese nuclear reactors were running as normal, however many were switched off in 2011. 
		Some reactors began to be returned to operation since 2010, so 2017 is taken to represent the lowest flux which is likely in the future, and 2010 the highest. 
		\added{Two reactors at Takahama were restarted in May and June 2017, so the average July-December was used for 2017.}
		If Japanese nuclear reactor activity returns to 2010 levels, then reactor neutrinos will be an important background to the DC channel of this analysis.

		Fast neutrons from cosmic ray muons are not a concern. 
		Neutrons produced by muons not passing through the detector should be few in number; fast neutrons will mostly not penetrate to the FV before capturing as they would need to pass through \textgreater4.5~m of water to do so. 
		Cosmic ray muons passing through the detector are detected very efficiently; products of spallation are already rejected at SK by vetoing events associated in time and space with a muon track (\cite{PhysRevD.93.012004}).
		Backgrounds from fast neutrons could be controlled by simply vetoing for 1~ms after each muon, which would introduce negligible dead-time.
		
		A fraction of muons create unstable daughter nuclei through spallation, and those that do can be efficiently identified through their light emission profiles(\cite{PhysRevD.91.105005,Li:2015lxa}). 
		Some unstable isotopes produce $\mathrm{\beta}$-delayed neutrons, which can form a DC candidate.
		Especially the $\mathrm{\beta n}$ decay of nitrogen-17 has a $\mathrm{\beta}$-ray energy in the energy range of interest. 
		This isotope should be efficiently identified by new spallation reduction methods, so 10\% dead-time and 95\% reduction is assumed, making it a small compared to other backgrounds.

		Reactor and geo neutrino IBD, spallation daughters decaying by $\beta$-delayed neutron, and neutrons from SF and ($\alpha,n$) are all evaluated with MC and added to the backgrounds calculated from pure water data. Remaining backgrounds after cuts are listed in \autoref{Sec:Selection}.

	\subsection{Event Selection}\label{Sec:Selection}

		The rate at which WIT recorded data in SK-IV was typically on the order of $10^8$ event candidates per day. 
		Most low energy triggers have a reconstructed position near the detector wall, and follow an exponentially falling energy and hit count distribution. 
		
		Initial cuts on the reconstructed vertex location and number of hit PMTs are used to select neutron candidates for the single neutron channel, reducing the background rate by a factor of $10^4$. 
		These cuts are based on the number of on-time hits, quality of reconstruction, and reconstructed vertex location. 
		This is 47\% efficient for simulated TNC on Gd candidates within the ID.
				The initial selection was based on the reconstructed vertex location and number of selected PMT hits, and the distance from the reconstructed vertex to the detector wall in the reconstructed direction of the event.

		DC events are selected by searching for events close in time that have been reconstructed within the FV.
		The time between the neutron and positron, and the distance between their reconstructed positions, is shown in \autoref{Fig:coiTimDist} for DC signal MC with 3~MeV positrons, and accidental background. 
		Other than the standard FV cut, selection of positron candidates is kept very inclusive, in order to achieve the lowest possible neutrino energy threshold. 
		The efficiencies of these cuts are somewhat energy dependent due to the deterioration of vertex resolution at low energy.
		A cut is also applied to delayed event candidates based on the distance from the reconstructed vertex to the detector wall, and the number of selected PMT hits.
		A boosted decision tree (BDT) classifier based on the angular distribution of hits, distance from the detector wall, reconstructed energy, and reconstruction quality is used to select good neutron candidates for both the DC and single neutron channels.
		The BDT was trained on a random 5\% sub-sample of the WIT background data, with performance evaluated on the whole dataset.
		Although the details of the Gd \gammaray{} cascade models may be refined when in-situ measurements are made using neutron sources, it is expected that the comparison between the two cascade models used is sufficient to cover potential differences in performance.
		
		\added{BDTs are a multivariate method capable of combining multiple discriminating variables in to a single score. 
		Cut criteria (decision trees) are automatically generated to separate signal and background training samples.
		A weight is assigned to each tree according to the accuracy of the tree.
		Events which are poorly classified are given boosted weight in subsequent trees.
		The overall score comes from the weighted sum of the response of all decision trees in the ensemble. 
		The TMVA (\cite{Hocker:2007ht}) software package was used.
		For a more detailed, rigorous explanation see \cite{Hocker:2007ht,FREUND1997119}.}

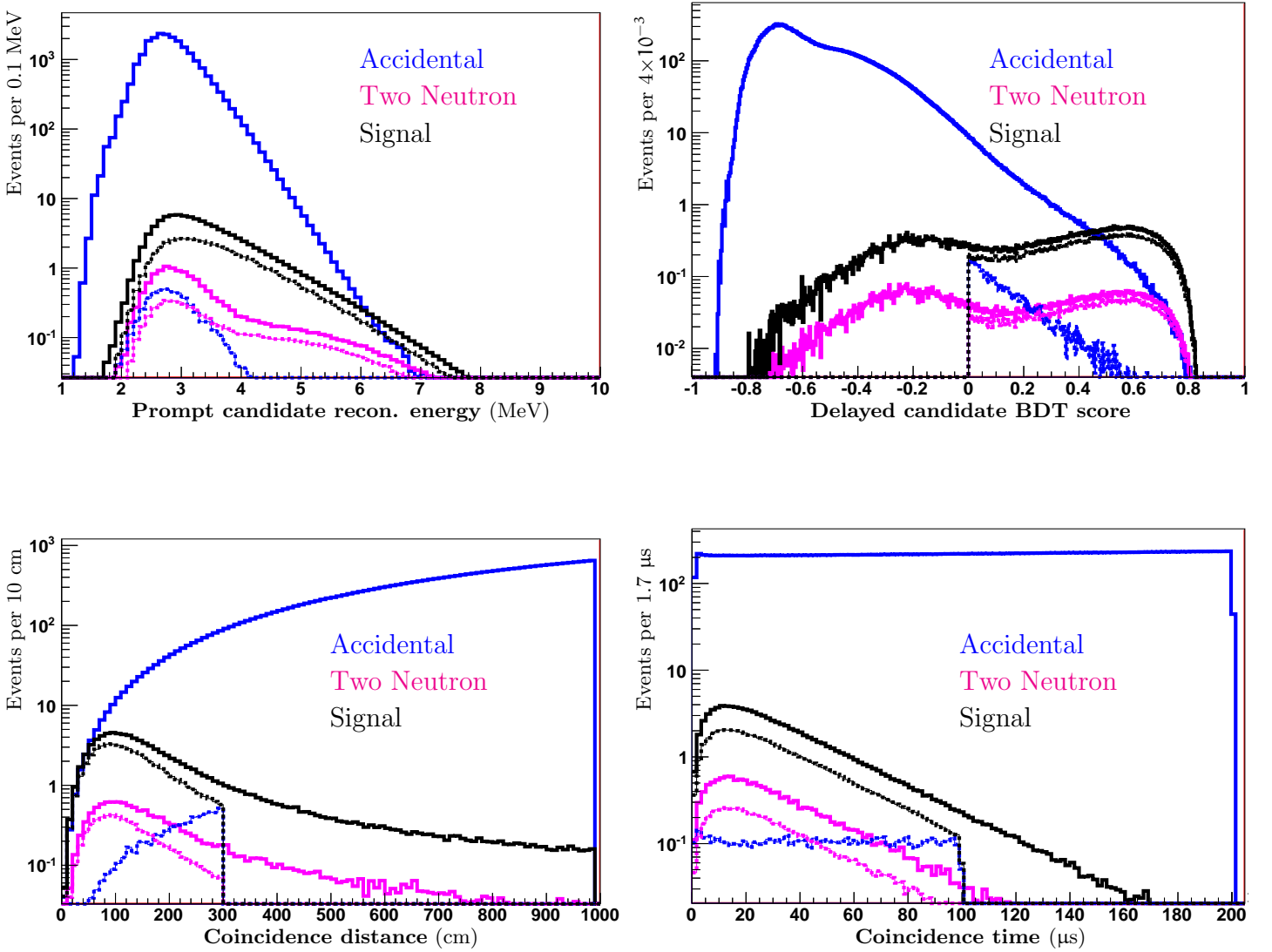
\begin{figure}
			\centering
			\gridline{	
												\begin{overpic}[trim={1cm 0.75cm 1.45cm 0},clip,width=0.5\textwidth]{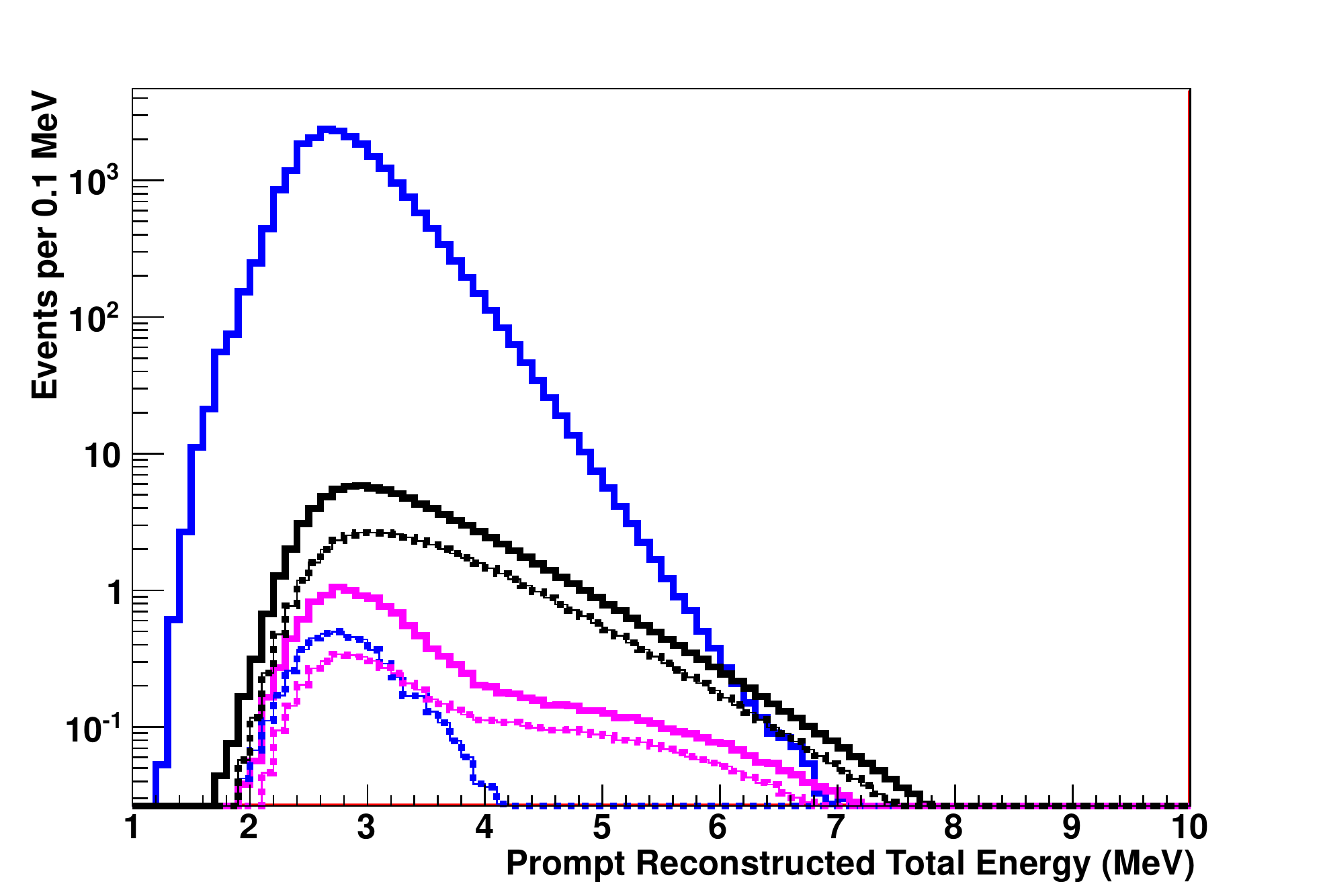}
					\put(45,50){\large					\begin{tabular}{l}
						\color{blue} Accidental\\
						\color{magenta} Two Neutron\\
						\color{black} Signal
					\end{tabular}}
					\put(50,-2){\makebox(0,0){\begin{tabular}{c}\textbf{Prompt candidate recon. energy} (MeV)\end{tabular}}}
					\put(-2,50){\makebox(0,0){\rotatebox[origin=c]{90}{Events per 0.1 MeV}}}
				\end{overpic}
				\hspace{0.3cm}
				\begin{overpic}[trim={1.0cm 0.75cm 1.9cm 0},clip,width=0.5\textwidth]{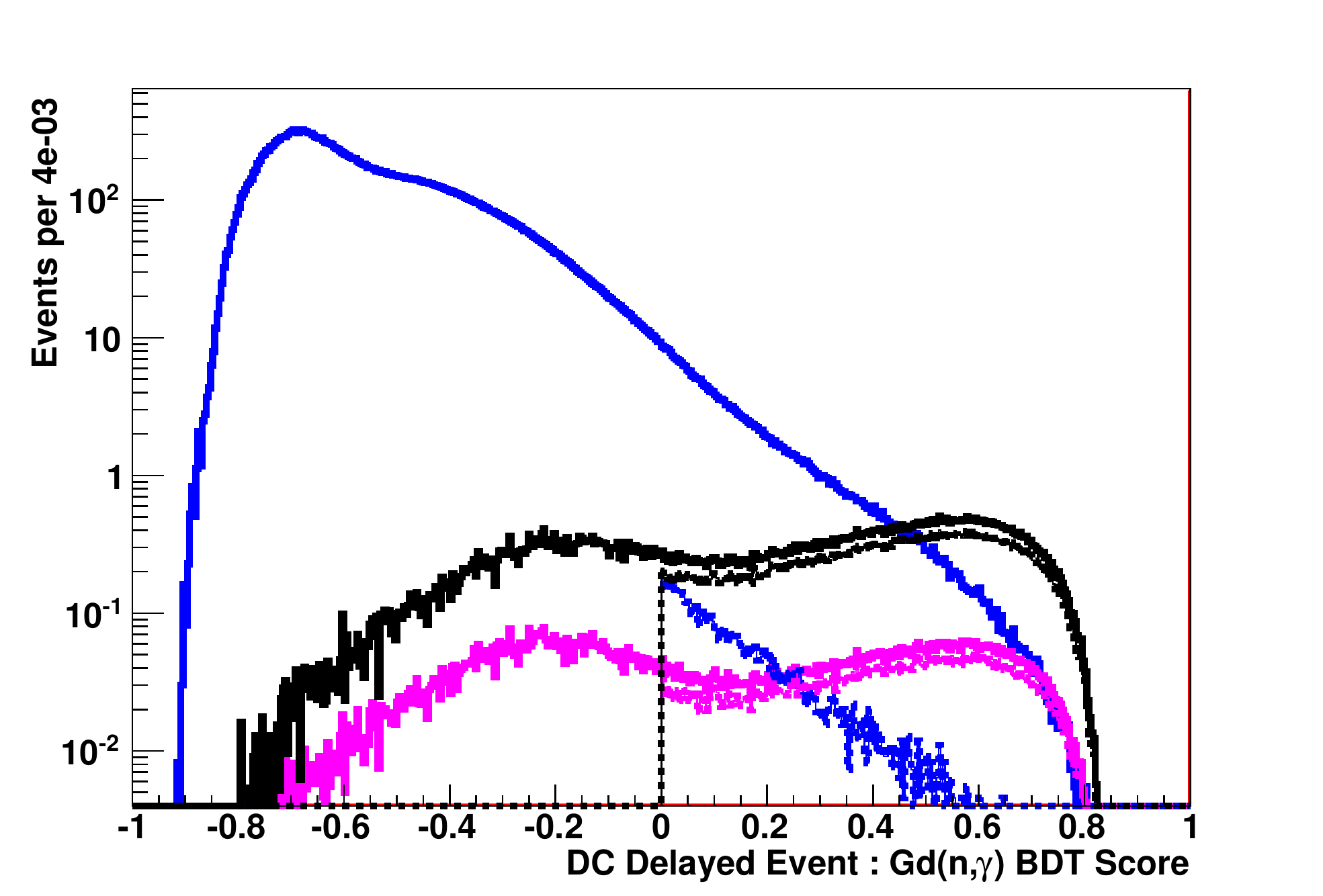}
															\put(45,50){\large					\begin{tabular}{l}
						\color{blue} Accidental\\
						\color{magenta} Two Neutron\\
						\color{black} Signal
					\end{tabular}}
					\put(50,-2){\makebox(0,0){\begin{tabular}{c}\textbf{Delayed candidate BDT score}\end{tabular}}}
					\put(-2,50){\makebox(0,0){\rotatebox[origin=c]{90}{Events per 4$\times10^{-3}$}}}
				\end{overpic}
			}
			\gridline{\vspace{0.1cm}}
			\gridline{
												\begin{overpic}[trim={1.0cm 0.75cm 1.45cm 0},clip,width=0.5\textwidth]{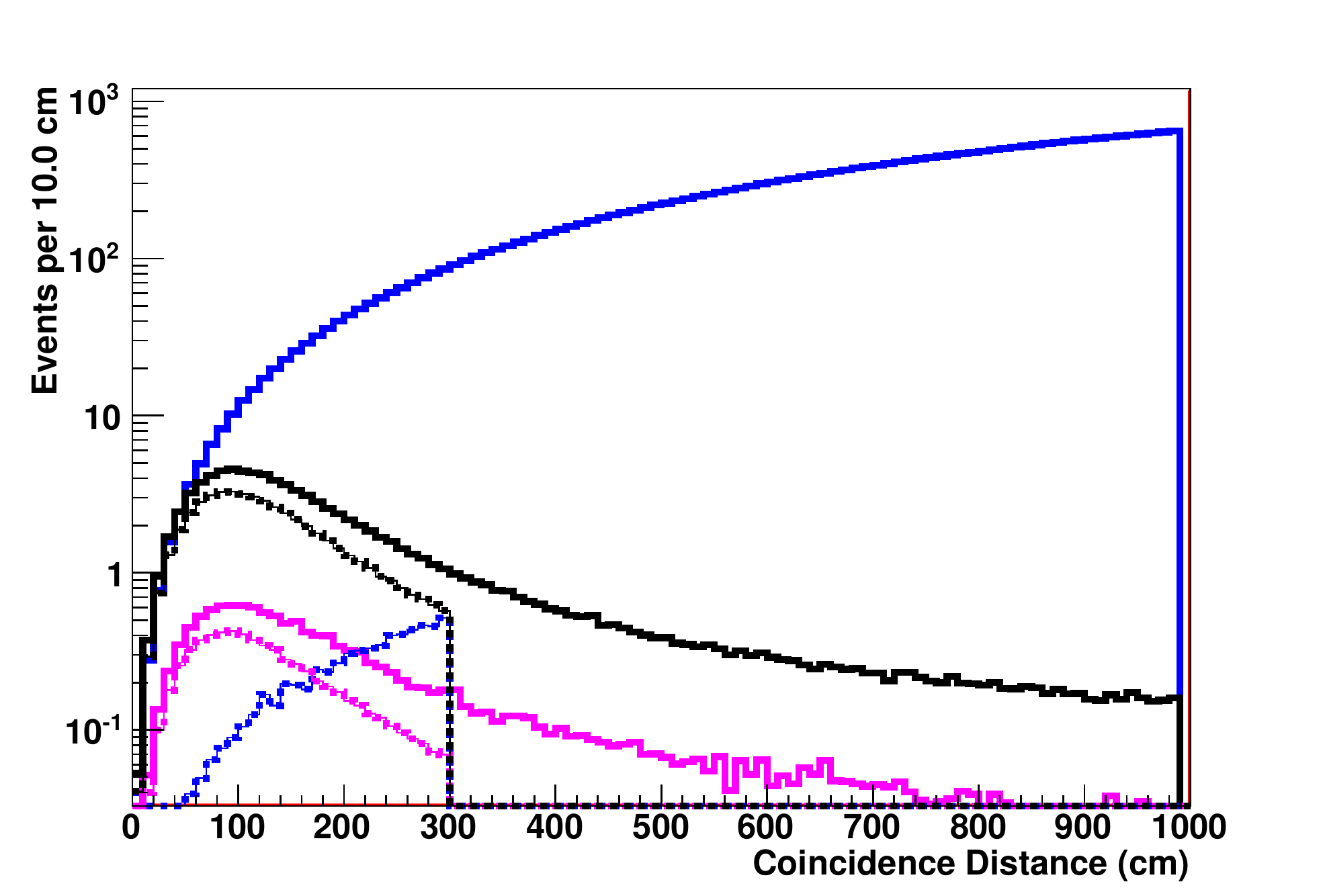}
															\put(40,40){\large					\begin{tabular}{l}
						\color{blue} Accidental\\
						\color{magenta} Two Neutron\\
						\color{black} Signal
					\end{tabular}}
					\put(50,-2){\makebox(0,0){\begin{tabular}{c}\textbf{Coincidence distance} (cm)\end{tabular}}}
					\put(-2,50){\makebox(0,0){\rotatebox[origin=c]{90}{Events per 10 cm}}}
				\end{overpic}
				\hspace{0.3cm}
				\begin{overpic}[trim={1.0cm 0.75cm 1.9cm 0},clip,width=0.5\textwidth]{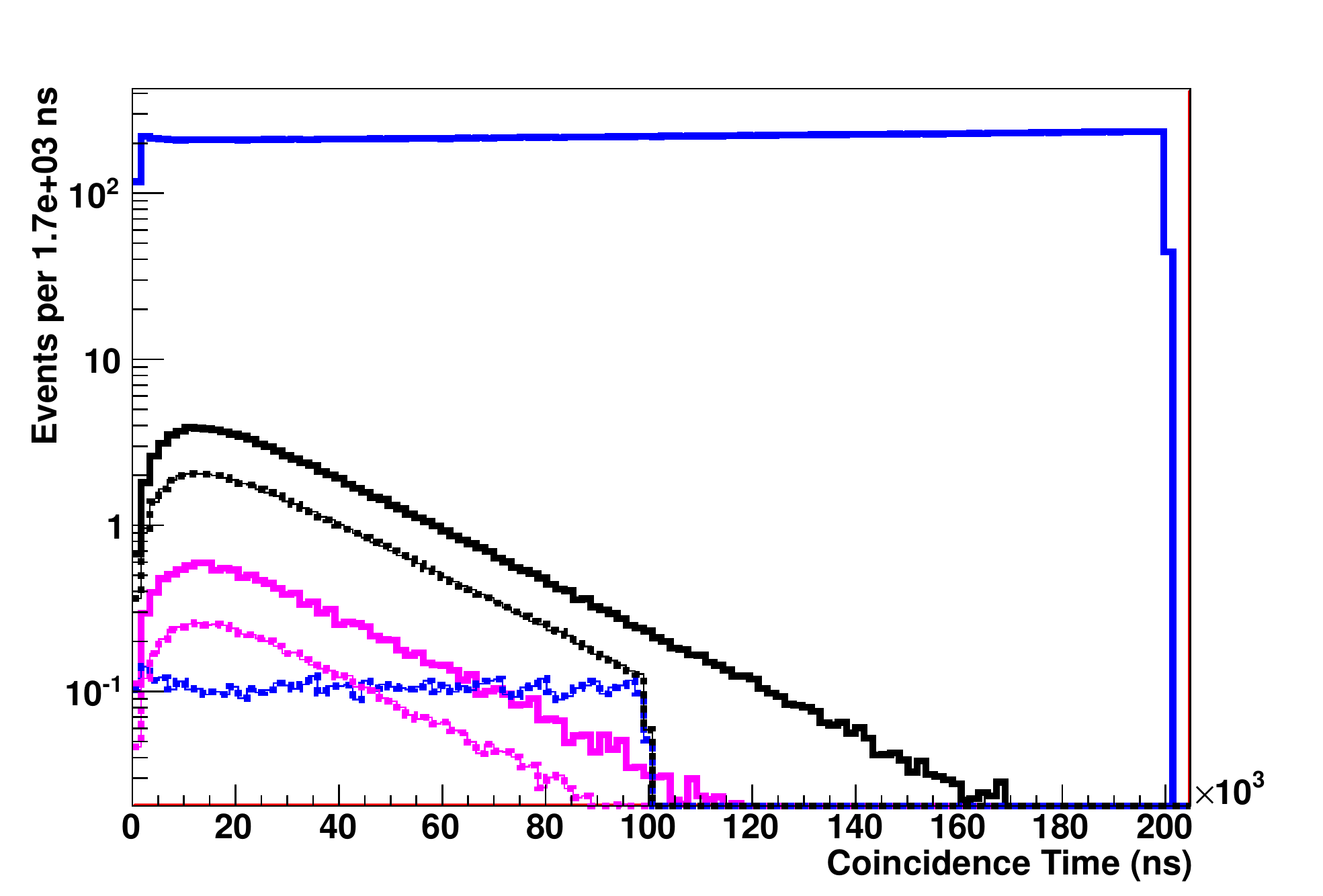}
															\put(40,40){\large					\begin{tabular}{l}
						\color{blue} Accidental\\
						\color{magenta} Two Neutron\\
						\color{black} Signal
					\end{tabular}}
					\put(50,-2){\makebox(0,0){\begin{tabular}{c}\textbf{Coincidence time }($\mathrm{\mu s}$)\end{tabular}}}
					\put(-2,50){\makebox(0,0){\rotatebox[origin=c]{90}{Events per 1.7 $\mu$s}}}
				\end{overpic}
			}
			\gridline{\vspace{0.1cm}}
			\caption{Distribution of DC channel variables in final selection for signal and largest backgrounds. Solid lines show before selection, dotted after.
				Background is normalised to 12~hours. Signal is normalised to the flux in the final 12~hours before CCSN for a 20~\Msun{} star at 500~pc.
				}
			\label{Fig:final_selection}
		\end{figure}

\begin{figure}\centering
																					\begin{overpic}[trim={1.0cm 0.72cm 1.45cm 0},clip,width=0.5\textwidth]{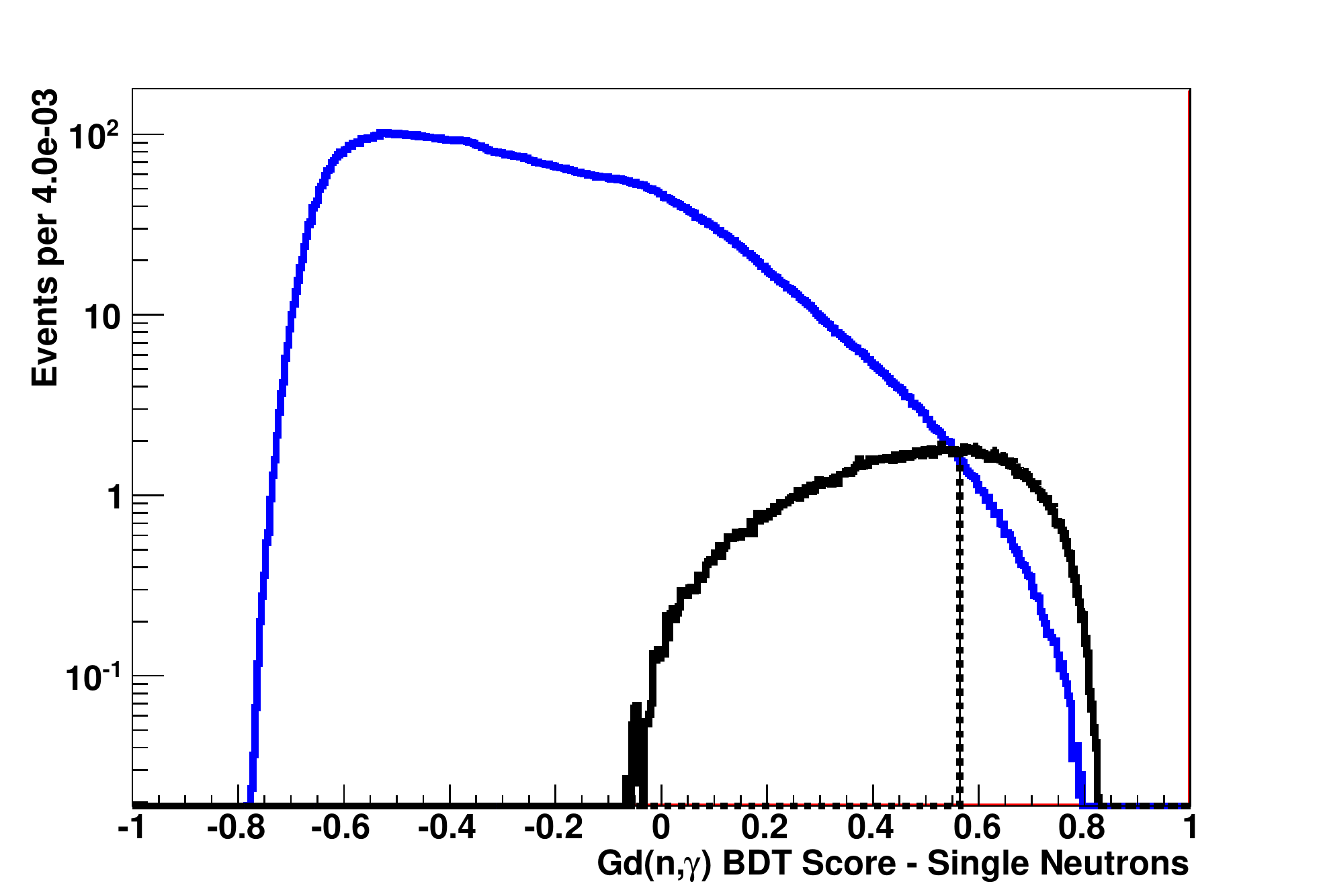}
				\put(55,54){\large					\begin{tabular}{l}
						\color{blue} Fake neutrons\\
						\color{black} Signal
					\end{tabular}}
					\put(50,-2){\makebox(0,0){\begin{tabular}{c}\textbf{Single neutron candidate BDT score}\end{tabular}}}
					\put(-2,50){\makebox(0,0){\rotatebox[origin=c]{90}{Events per 4$\times10^{-3}$}}}
				\end{overpic}
				\gridline{\vspace{0.1cm}}
						\caption{
				Distribution of single neutron channel Gd BDT score in final selection for signal and largest background. Solid lines show the distribution before selection, the dotted line shows the cut point.
				Background is normalised to 12~hours. Signal is normalised to the flux in the final 12~hours before CCSN for a 20~\Msun{} star at 500~pc.
				The distribution of the Gd BDT score is different from that shown in \autoref{Fig:final_selection} as single neutron events had a higher hit threshold applied in pre-selection.
				}
			\label{Fig:final_selection_sngl}
		\end{figure}

		Cuts were optimised to give the greatest range for efficient detection of a 25~\Msun{} pre-SN star using the model of \cite{OdrzywolekHeger} 0.1~hour before core collapse, assuming the more optimistic background and efficiency cases. 				The distribution of the variables used in the final selection of DC events are shown in \autoref{Fig:final_selection} for the two largest backgrounds in the most important variables.
		While the accidental backgrounds are well controlled by the combination of coincidence variables and the neutron BDT, the background from pairs of neutrons is harder to reduce. 
				Selection of single neutrons was based mostly on the BDT classifier, with distributions in the final selection shown in \autoref{Fig:final_selection_sngl}. 
		\added{The cuts shown in \autoref{Fig:final_selection} are: delayed candidate BDT score $>0$, coincidence distance $<300~cm$, coincidence time $<100~\mu s$. The cut shown in \autoref{Fig:final_selection_sngl} is: candidate BDT score $>0.56$.}\explain{ Reviewer comment: + Fig 6. final selection criteria should be clearly written. I guess coincidence distance $<$ 300 cm, coincidence time $<$ 1000us and BDT score $>$ 1.  Are they correct ?}


		\explain{acceptance->efficiency, in response to reviewer comment: And, I would like to confirm the definition of acceptance.   So, the total detection efficiency is from the WIT trigger efficiency and signal acceptance, or total detection efficiency is same to the signal acceptance.}
		Total \replaced{signal acceptance}{trigger and selection efficiency} depends on the energy distribution of the flux.
		In the final 12~hours before core collapse, the proportion of all IBD events which are \replaced{accepted}{triggered and selected} is 
		4.3-6.7\% for DC events, and 9.5-10\% for neutron singles. 		Using the alternative MC for the \gammaray{}s~from TNC on gadolinium, these numbers are 3.9-6.1\% and 7.3-8.0\% respectively. 		\added{A range is given as these are flux averaged efficiencies, and therefore depend on the flux spectrum.}
		\explain{Reviewer comment:+ What is the largest uncertainty of the acceptance 4.3 - 6.7\% for DC events and 9.5-10\% for neutrino singles. Author response: the range of acceptances given here are to account for the energy distribution of the flux. I agree that the way it was written was confusing. I hope by including a plot of efficiency against energy this is resolved.}
		\added{\autoref{Fig:Eff:Final} shows the total efficiency for DC events against the true positron energy. 
		The efficiency for selecting TNC on Gd does not depend on the true positron energy.
		Note that this efficiency curve has been directly optimised for detecting the pre-SN flux in the final 12~hours, so should not be used in higher energy contexts.
		\begin{figure}
			\center
			\includegraphics[width=0.49\textwidth]{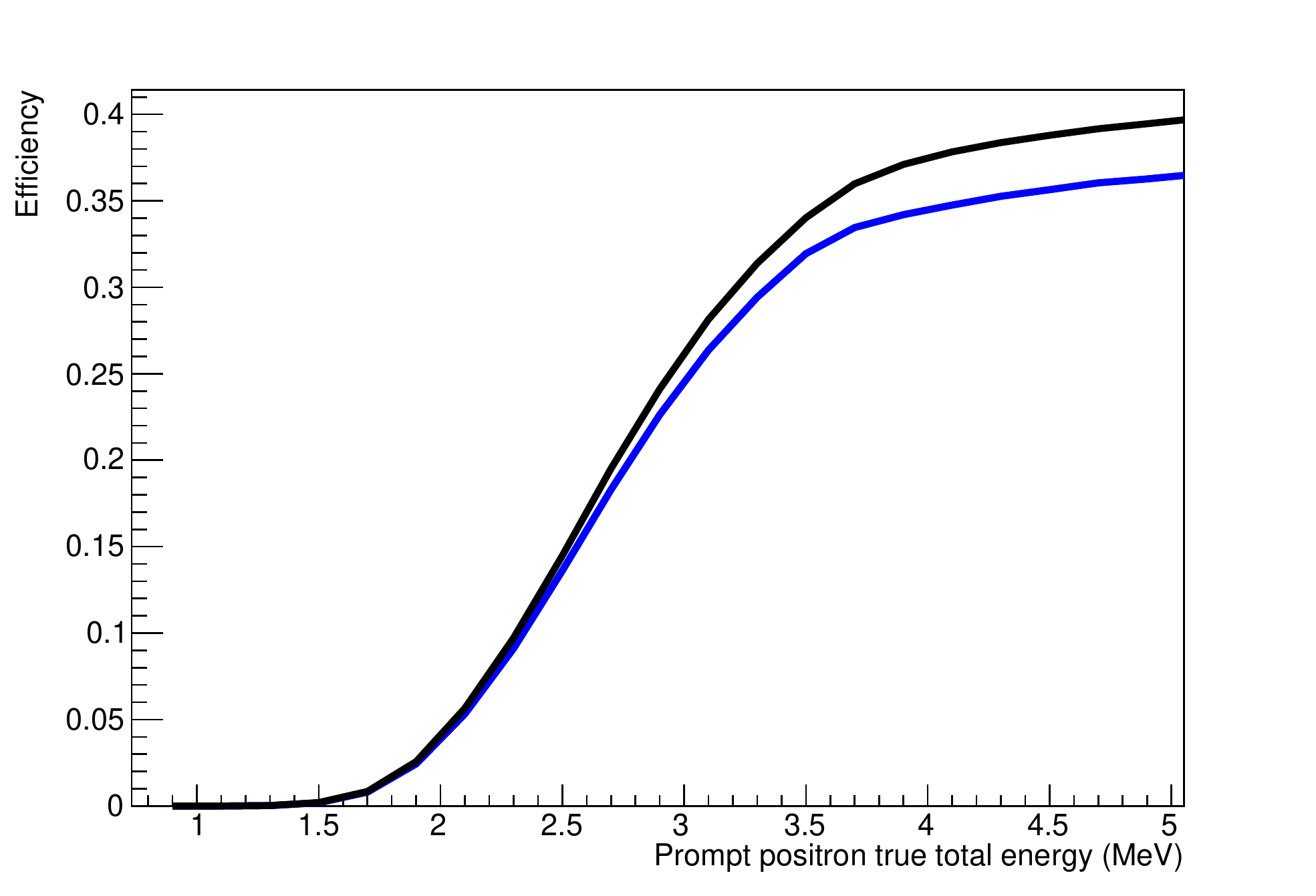}
			\caption{
				Trigger and selection efficiency for DC events against prompt positron true total energy. 
				Above 4~MeV positrons are efficiently triggered and reconstructed, and so the efficiency is determined by the triggering and selection of the delayed part of the event.
				The result with the alternative Gd \gammaray{}-cascade model is shown in \color{blue}{blue}.
			}
			\label{Fig:Eff:Final}
		\end{figure}}

				The selection requirements on the DC time, and the positron quality requirements, are very efficient for signal.
		DC signal efficiency is lower at low energies due to the lower trigger efficiencies, and worse vertex reconstruction leading to increased DC distance.
		It should not be alarming that this efficiency is lower and these backgrounds higher than are typically quoted for other SK analyses (such as the supernova relic neutrino analysis), which are at higher energy. 
		At higher energy, reconstruction is much better, and backgrounds much lower. 
		Almost all prompt events in this analysis reconstruct below the energy threshold of any other analysis at SK.
		If the TNCs were only on H, rather than Gd, there would be no sensitivity at all in this analysis.

		Remaining backgrounds for single neutrons, dominated by fake neutrons from the intrinsic radioactivity of the detector, total 66-140 per 12~hour window. 
		\replaced{For DC events, remaining backgrounds are 5 to 11 accidental, 0.2 to 0.4 from SF \gammaray{}s~in coincidence with neutrons, 6.8 to 14 from pairs of neutrons, 0.2 to 3.0 from reactors; in total 12 to 28 per 12~hour window.}{For DC events, remaining backgrounds are 5 to 11 accidental, 0.2 to 0.4 from SF \gammaray{}s~in coincidence with neutrons, 6.8 to 14 from pairs of neutrons, 0.3 to 3.0 from reactors; in total 12 to 28 per 12~hour window.}\explain{Reviewer comment: In 2017, some Takahama reactors restarted. Authors should use the data clearly before or after Tahakama restart. Author response: we changed to using July onwards only for 2017, which increases the background in the low reactor case slightly, though not enough to effect the bottom line result.}
		
	\subsection{Detection Strategy} \label{Sec:Strategy}

	The simplest way of searching for a rapid excursion in the candidate event rate is to define some signal time window (e.g. 12~hours), and background window (e.g. 30~days), then perform a hypothesis test. 
	The null hypothesis is that the observed rate in the signal window is consistent with the observed rate in the background window, taking into account Poissonian fluctuations.
	The alternative hypothesis is simply that the rate in the signal window is higher than that in the background window.
	A Poisson likelihood for the null hypothesis is calculated for the total detected event rate, combining both the DC and single neutron channels. 
	
	The time from the start of a data block for reconstruction, event selection, and hypothesis testing will be around 10 minutes. 	This sets an estimated typical latency for an alarm.
	
	The length of the signal window should be similar to the timescale over which the event rate would change if a pre-SN star was detected.
	The choice of signal window size can have a dramatic effect on the sensitivity of the analysis, as it affects the statistical fluctuations, the background level, and the trial factor.
	A range from 1 to 72 hours were tested and the 12~hour signal window performed best. 
	For the model used, the \tantinu{e} flux multiplied by the IBD cross section, integrated over the final 12~hours before collapse is summarised in \autoref{tab:flux}.
		
	\begin{table}
		\centering
		\begin{tabular}{ccc|c}
			MO & Mass & Model & \tantinu{e} flux in final 12 hours\\
			&(\Msun{})&& ($ \mathrm{10^{12} cm^{2}}$~per~12~hours~per~nucleon) \\
			\toprule
			NO & 15 & \cite{OdrzywolekHeger} & 1.6 \\
			 & 15 & \cite{Patton:2017neq} & 1.9 \\
			 & 25 & \cite{OdrzywolekHeger} & 3.3 \\
			 & 30 & \cite{Patton:2017neq} & 3.8 \\
			\midrule
			IO & 15 & \cite{OdrzywolekHeger} & 0.44 \\
			 & 15 & \cite{Patton:2017neq} & 0.53 \\
			 & 25 & \cite{OdrzywolekHeger} & 0.93 \\
			 & 30 & \cite{Patton:2017neq} & 1.2 \\
			\bottomrule
					\end{tabular}
		\caption{Anti-electron neutrino flux multiplied by IBD cross section for models considered.}
		\label{tab:flux}
	\end{table}

	A longer background window would always be better due to reduced uncertainty in the background rate, however the background rate may change slowly over time. 
	A gradual change might be expected, due to the gradual increase in PMT gain, seasonal variations in the radon concentration in the mine air 
	 (\cite{Pronost:Radon,Radon:NAKANO2017108}), changes in water flow affecting radon activity in the fiducial volume, or changes in nearby nuclear power station activity. 
	 In this study it is assumed that the background level is known precisely when the signal is detected.
		
	These methods are chosen for the purpose of benchmarking performance. 
	In practice, greater sensitivity can be achieved by properly accounting for the likelihood of the event rate across separate bins, and by assuming a more complicated alternative hypothesis, for example by calculating the rate of increase of the candidate event rate.
	Furthermore, a model including background time variation, measurement uncertainty, and correlation between bins, should be included when a large enough background sample has been collected during SK-Gd.

	In attempting to get a SN early warning from the detection of pre-SN neutrinos, there are four variables which together describe a detector's performance. 
	\begin{enumerate}
		\item Alarm efficiency, i.e. the probability of correctly detecting a true pre-SN
		\item False positive rate (FPR)
		\item Expected time of early warning that the detector would provide
		\item Expected distance to which the warning would be efficient
	\end{enumerate}
	Tolerating a higher FPR would allow for greater range and more warning, but would reduce trust in the warning system; a choice needs to be made on what FPR is acceptable. 
	Note that a SN within 1~kpc is a rare occurrence - on the order of 1 in 10,000 years based on historical data (\cite{Adams:2013ana}). 
	FPR levels shown in this paper are 1~per~year and 1~per~century~(cy.), and are assumed to be set by Poisson fluctuations only.
	Range is defined as the point at which alarm efficiency is 50\%.
	By formulating the problem in terms of FPR, a trials factor is incorporated.

\section{Results}\label{Sec:results}
\begin{figure}[htbp]
		\gridline{
			\fig{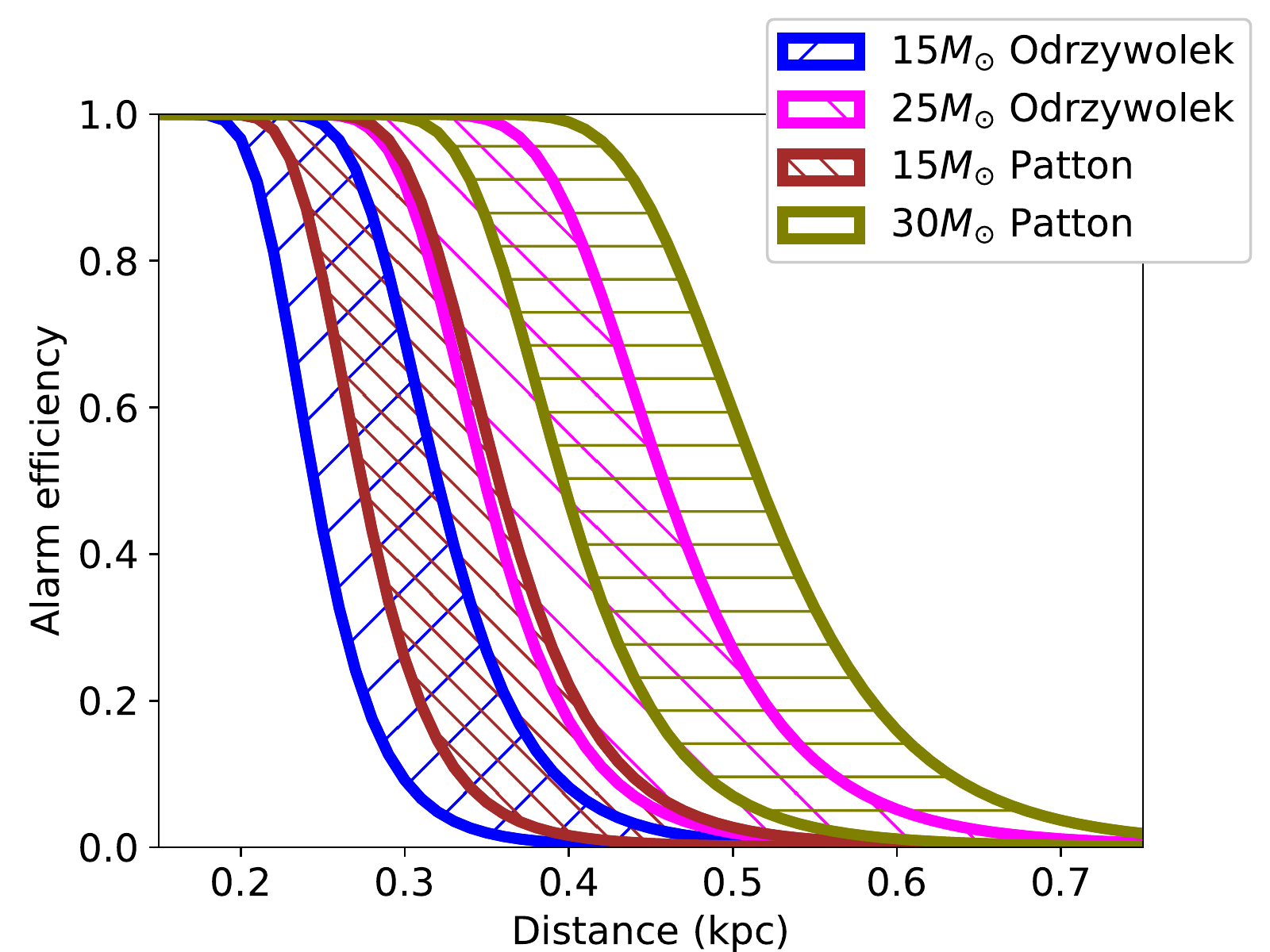}{0.49\textwidth}{(a) NO, FPR=1/cy.}
			\fig{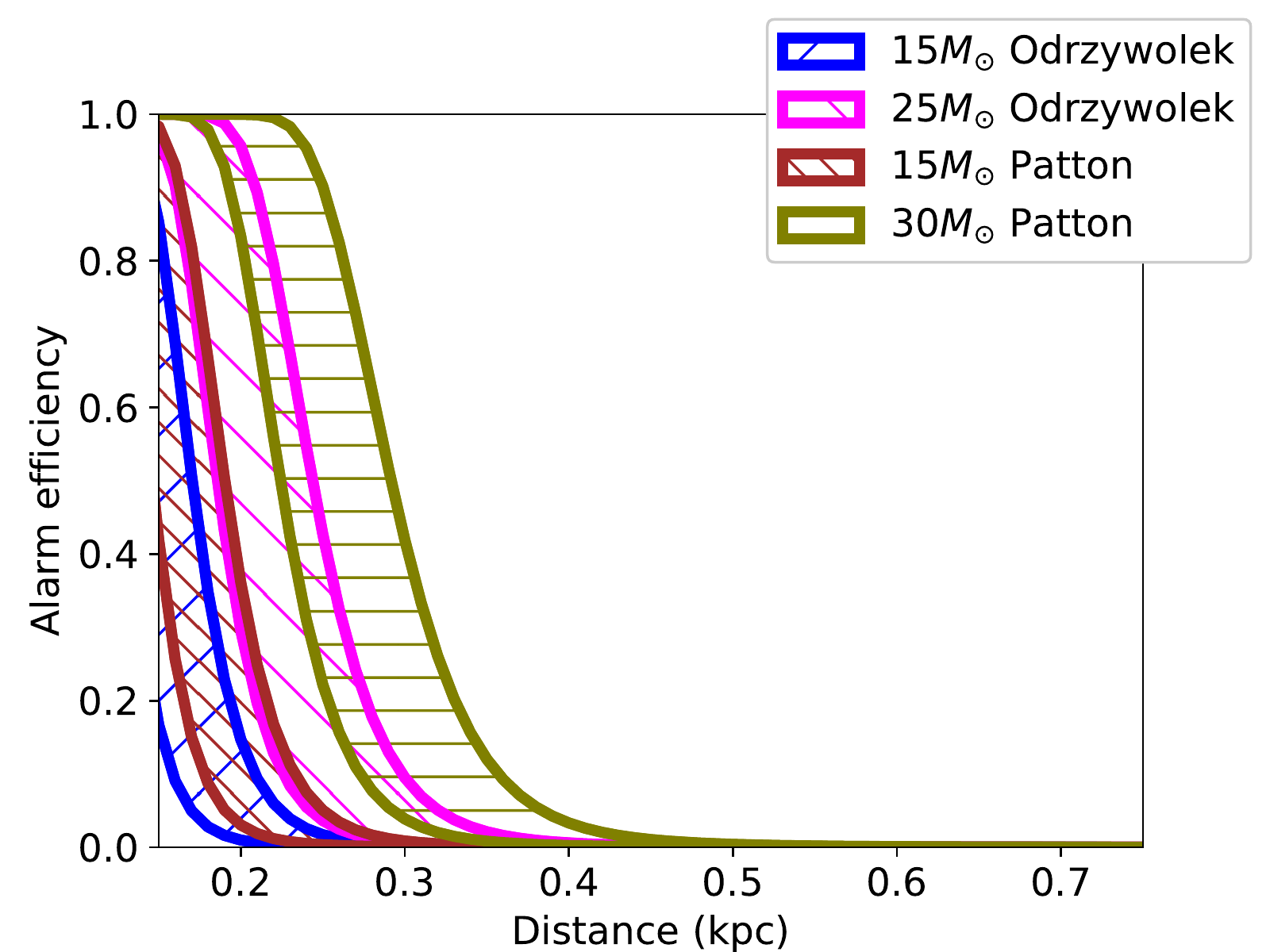}{0.49\textwidth}{(b) IO, FPR=1/cy.}
		}
		\gridline{
			\fig{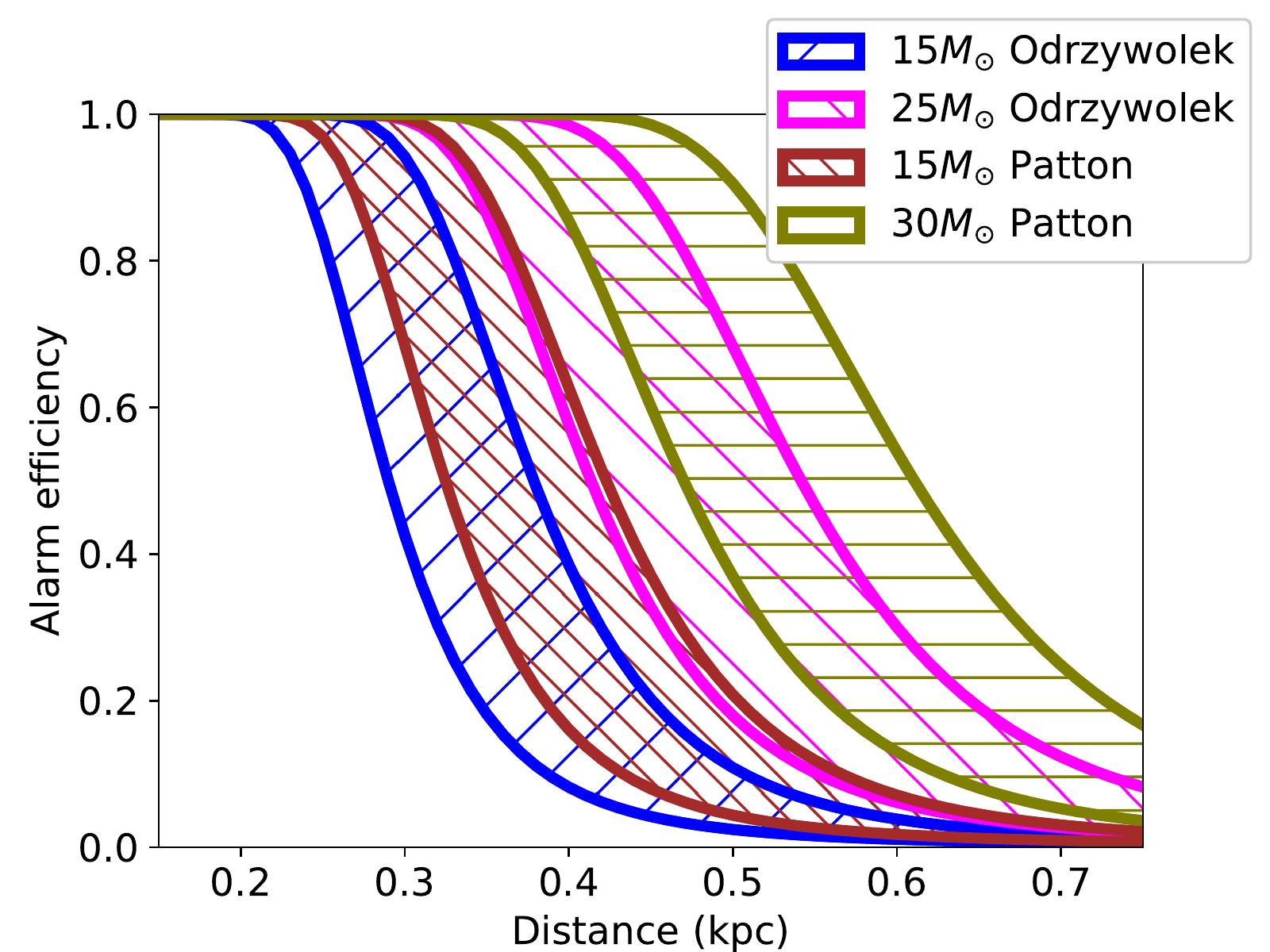}{0.49\textwidth}{(c) NO, FPR=1/year}
			\fig{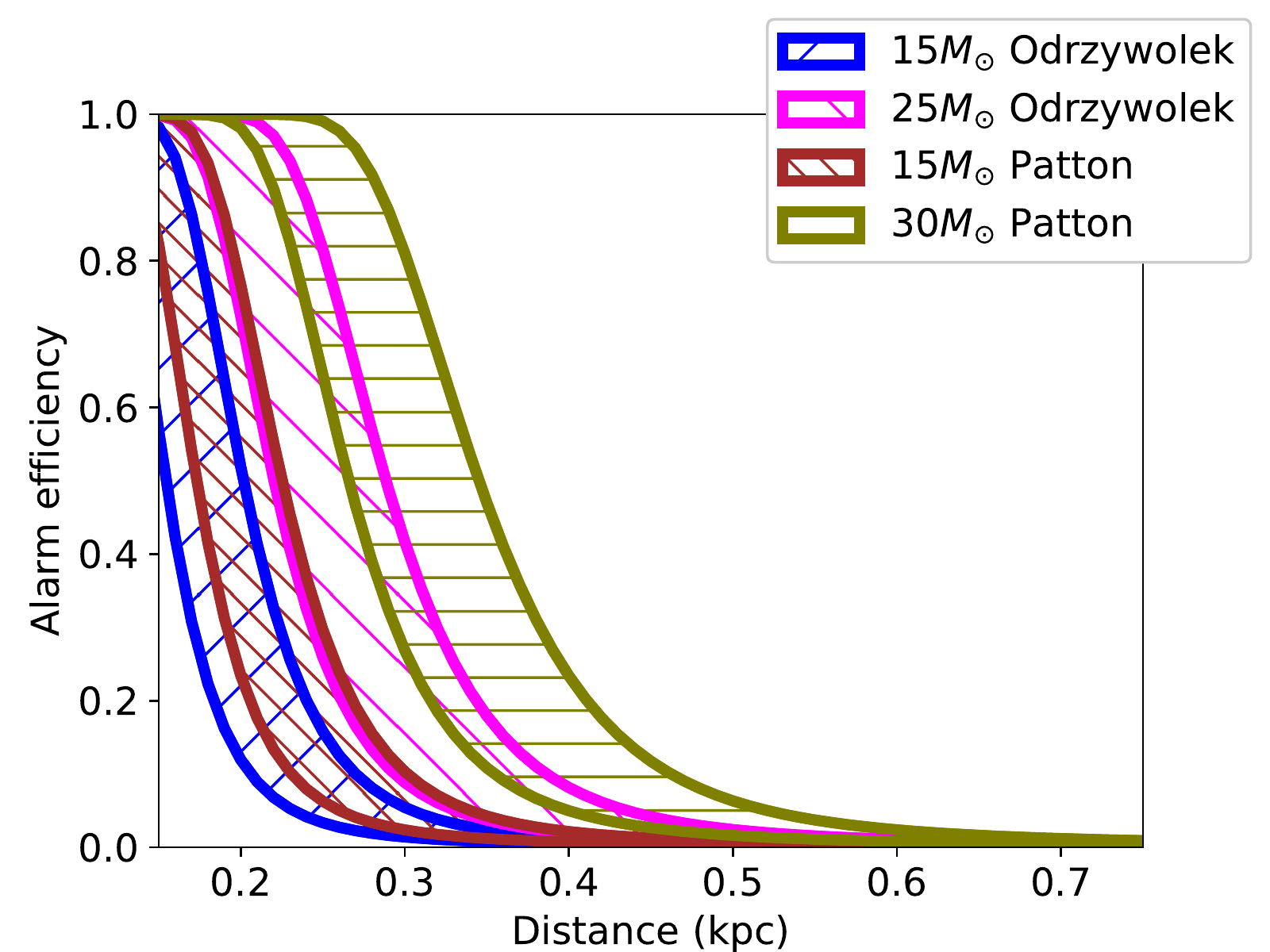}{0.49\textwidth}{(d) IO, FPR=1/year}
		}
		\caption{Expected maximum range of detection in the final 12 hours before collapse, for 15~\Msun{} and 25~\Msun{} stars.
			The width of the bands shows uncertainty due to levels of background in SK-Gd, and the difference between models of \gammaray{} emission from TNC on Gd.
		}
		\label{Fig:RangePatton}
	\end{figure}

\begin{figure}[p]
		\gridline{
			\fig{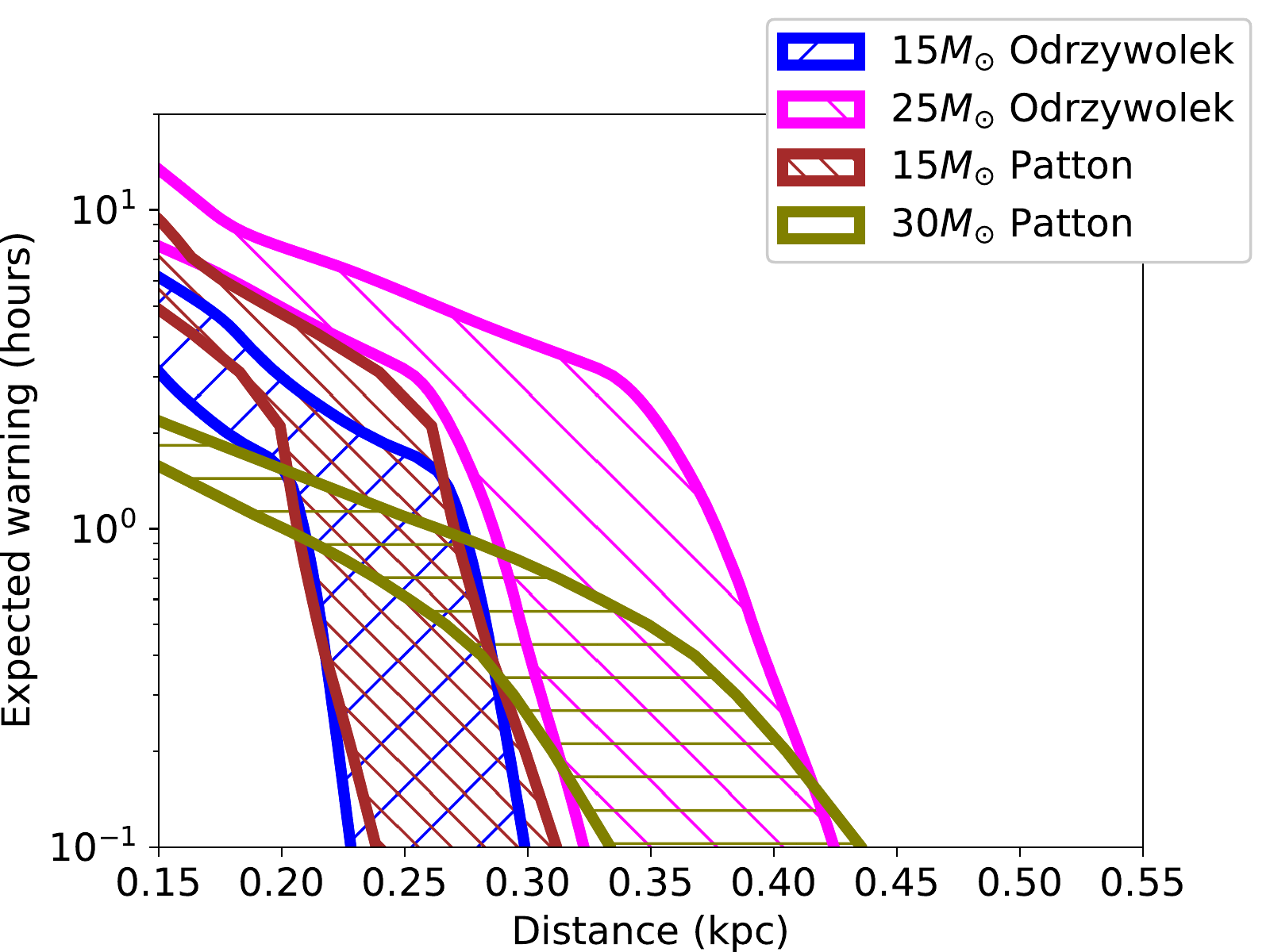}{0.49\textwidth}{(a) NO, FPR=1/cy.}
			\fig{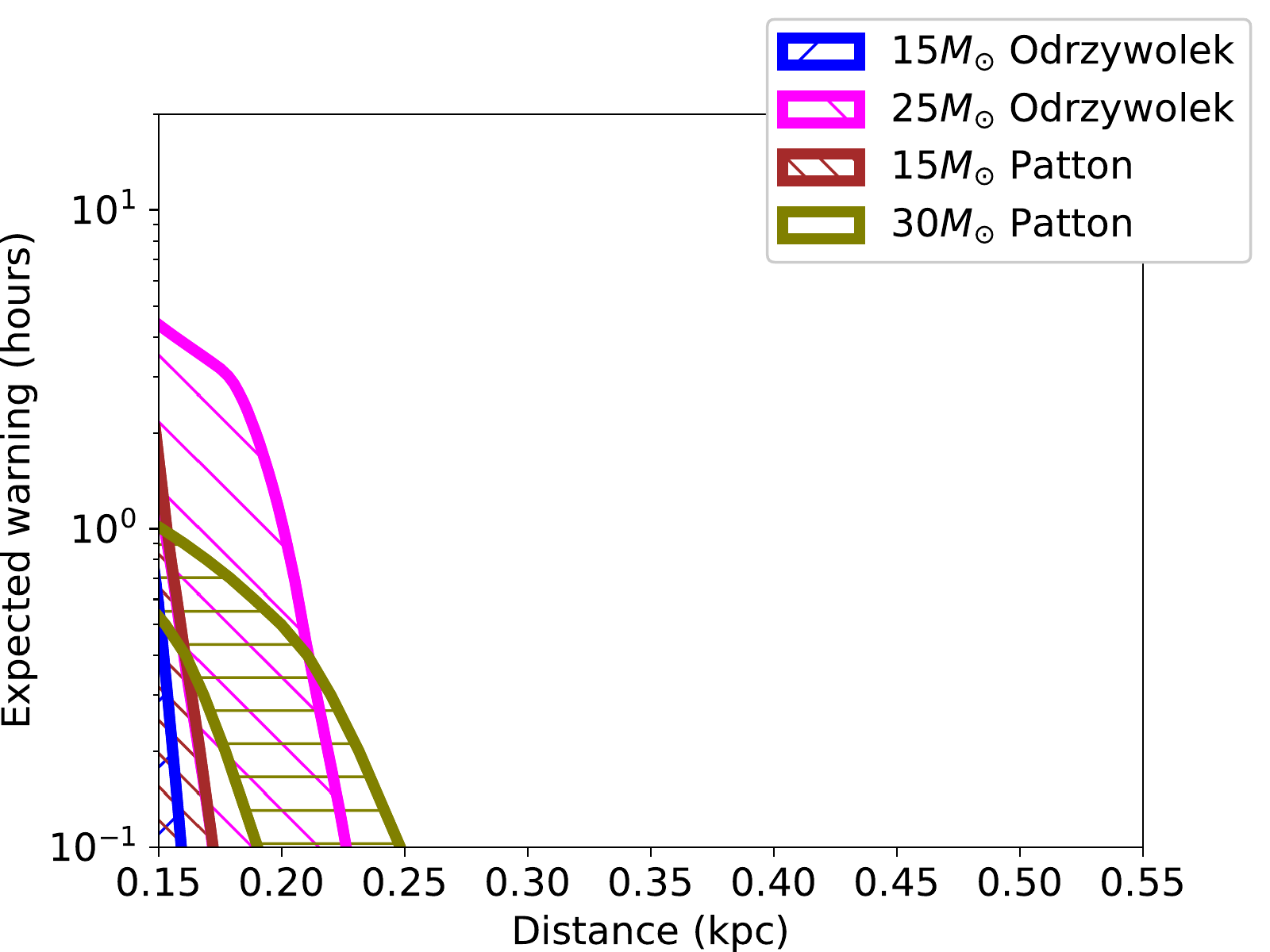}{0.49\textwidth}{(b) IO, FPR=1/cy.}
		}
		\gridline{
			\fig{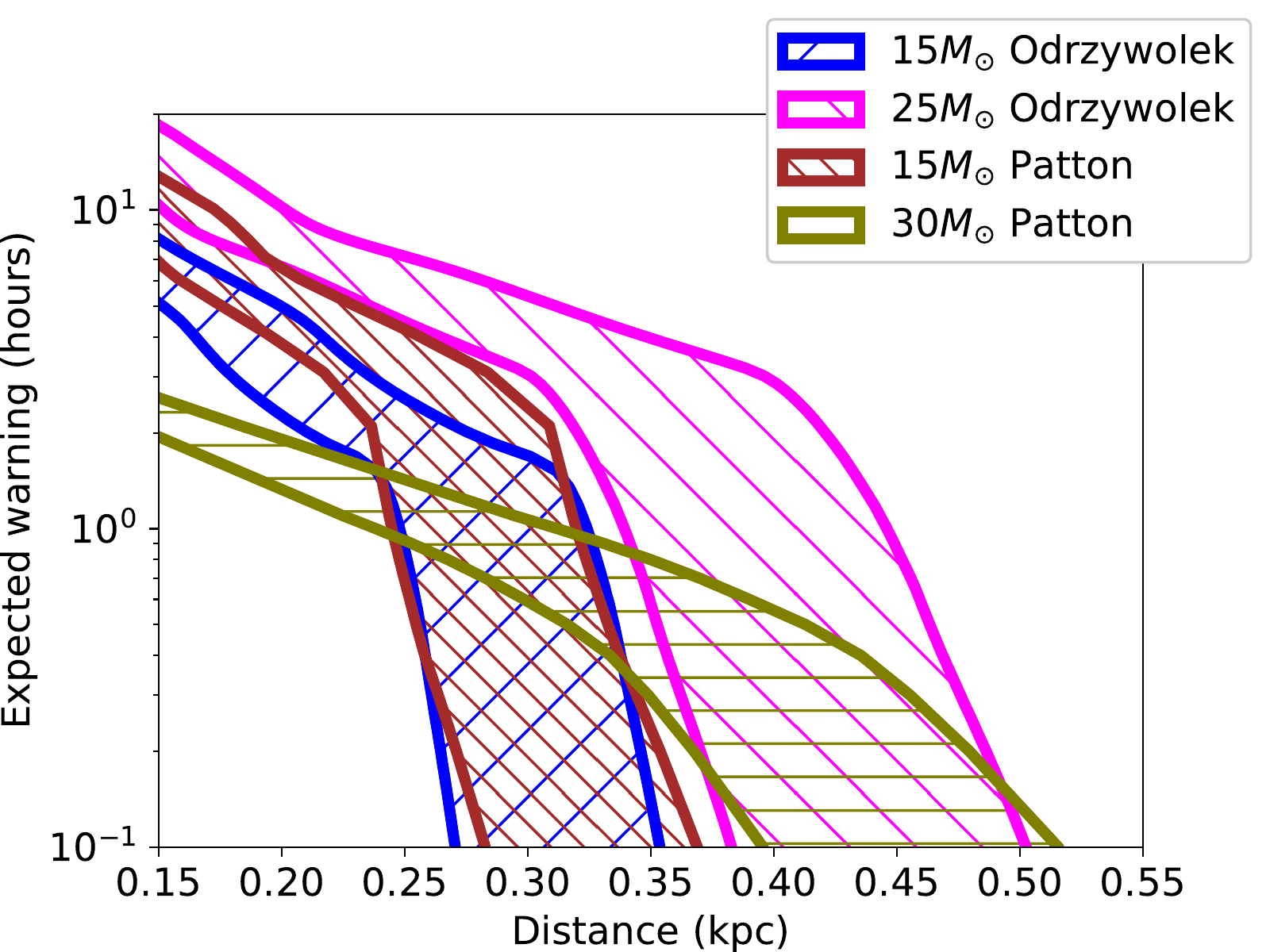}{0.49\textwidth}{(c) NO, FPR=1/year}
			\fig{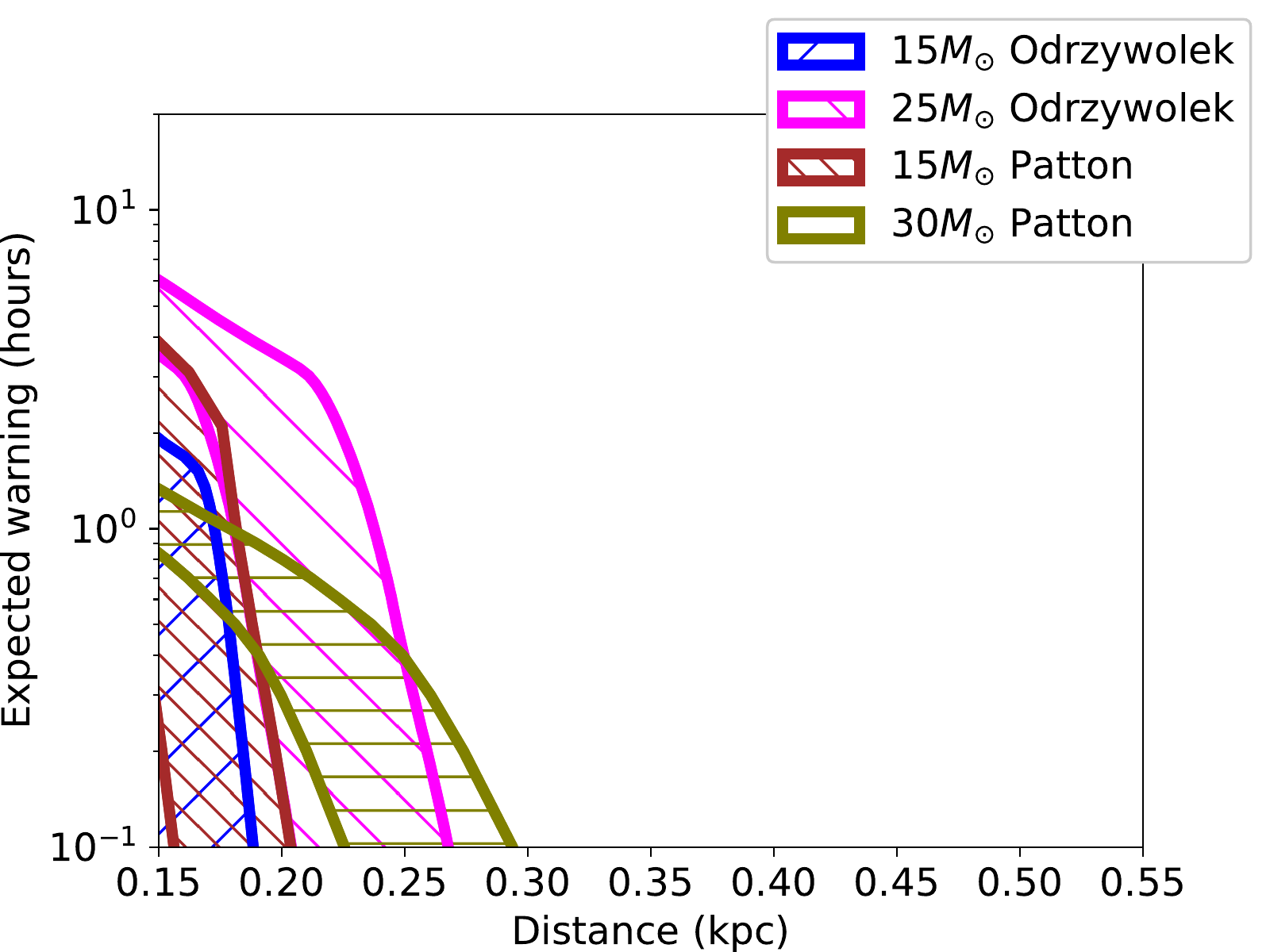}{0.49\textwidth}{(d) IO, FPR=1/year}
		}
		\caption{
			Expected early warning against range of detection, for 15~\Msun{} and 25~\Msun{} stars. 
			It is assumed the range is the largest distance at which probability of detection is greater than 50\%.
		}
		\label{Fig:WarningPatton}
	\end{figure}

	\autoref{Fig:RangePatton} shows the distance to the pre-SN star at which the null hypothesis would be rejected before core collapse. 
	\autoref{Fig:WarningPatton} shows the largest amount of time before core collapse at which a pre-SN is expected to be detected.
	The width of the bands shows uncertainty due to levels of background and the difference between models of \gammaray{} emission from TNC on Gd.
	The distances at which alarm efficiency is above 50\% are summarised in \autoref{tab:range}.

		Results depend on the neutrino mass ordering (normal \emph{NO} or inverted \emph{IO}), the ZAMS mass of the star, the distance to the star, the background level in SK-Gd.
	Questions of detector model uncertainty will be resolved by in-situ measurements once SK-Gd is loaded. 
	An inverted neutrino mass ordering is detrimental to this analysis as it reduces the \tantinu{e} fraction of the pre-SN flux.

	\begin{table}
		\center
		\begin{tabular}{ccc|rcr|rcr}
		&&&\multicolumn{6}{c}{Max. range (pc)}\\
		&Mass&&\multicolumn{6}{c}{given FPR}\\
		   &(\Msun{}) & Model &  \multicolumn{3}{c}{1/year} &\multicolumn{3}{c}{1/cy.} \\
		   \toprule
  			NO	&	15 & Odrzywolek	& 300 &-&400 & 250 &-&300\\
  				&	15 & Patton	&	 420 &-&600 & 360 &-&500\\
				&	25 & Odrzywolek		& 330 &-&400 & 280 &-&400\\
				&	30 & Patton		& 480 &-&600 & 410 &-&500\\
		\midrule
		IO &	15 & Odrzywolek	&160 &-&200 & 130 &-&200 \\
			&	15 & Patton	&220 &-&300 & 190 &-&200 \\
			&	25 & Odrzywolek	&180 &-&200 & 150 &-&200 \\
			&	30 & Patton	&270 &-&400 & 230 &-&300 \\
			\end{tabular}
		\caption{
			Maximum range of detection under various assumptions. 
			Range is assumed to be the highest distance at which alarm efficiency is greater than 50\%.
			Uncertainty comes from TNC \gammaray{} model, and background uncertainty.
			The same information is contained in \autoref{Fig:RangePatton}.
			}
		\label{tab:range}
	\end{table}

\begin{figure}[htbp]
		\center
		\gridline{
			\fig{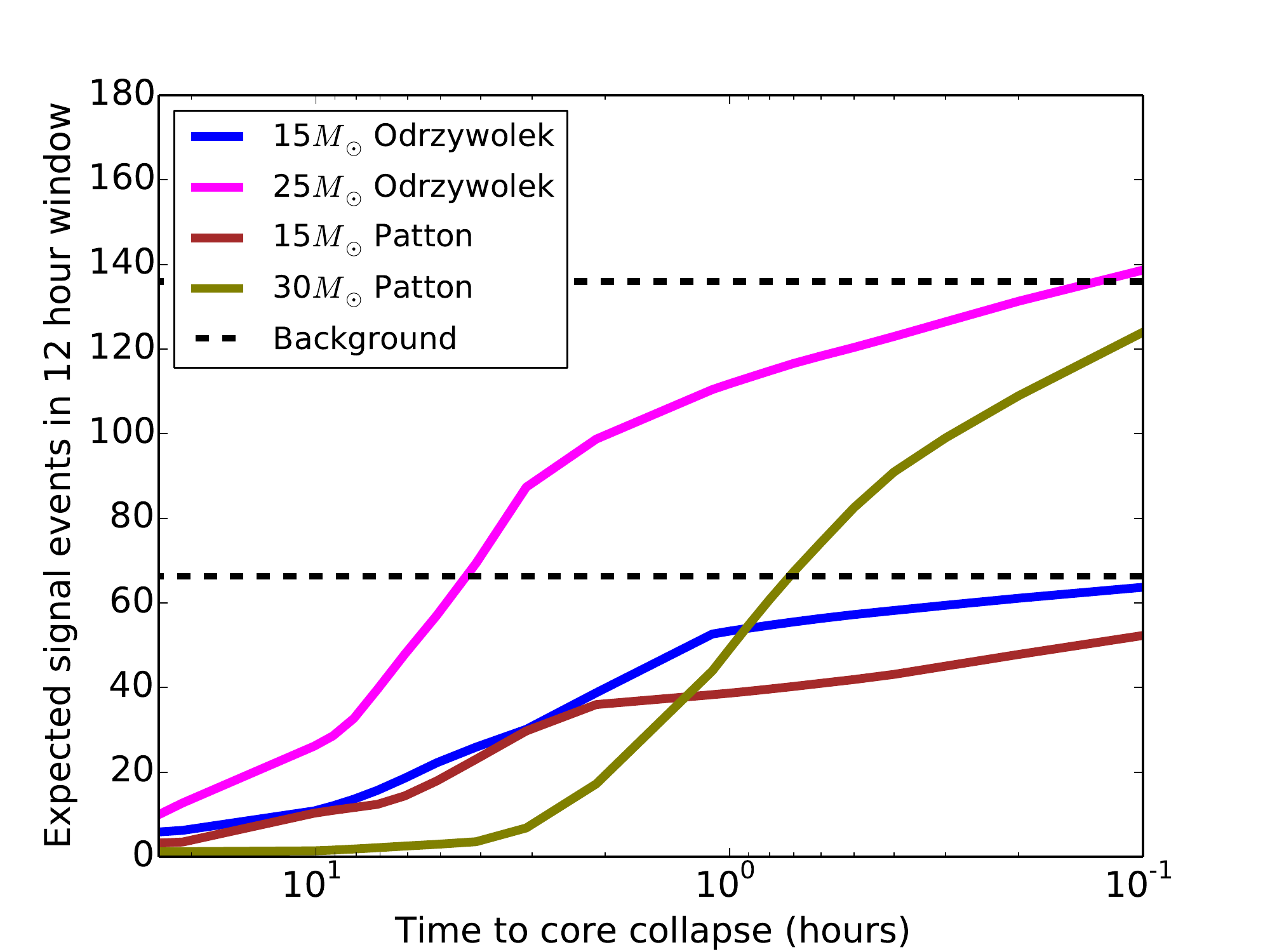}{0.49\textwidth}{(a) Neutron singles expected signal rate}
			\fig{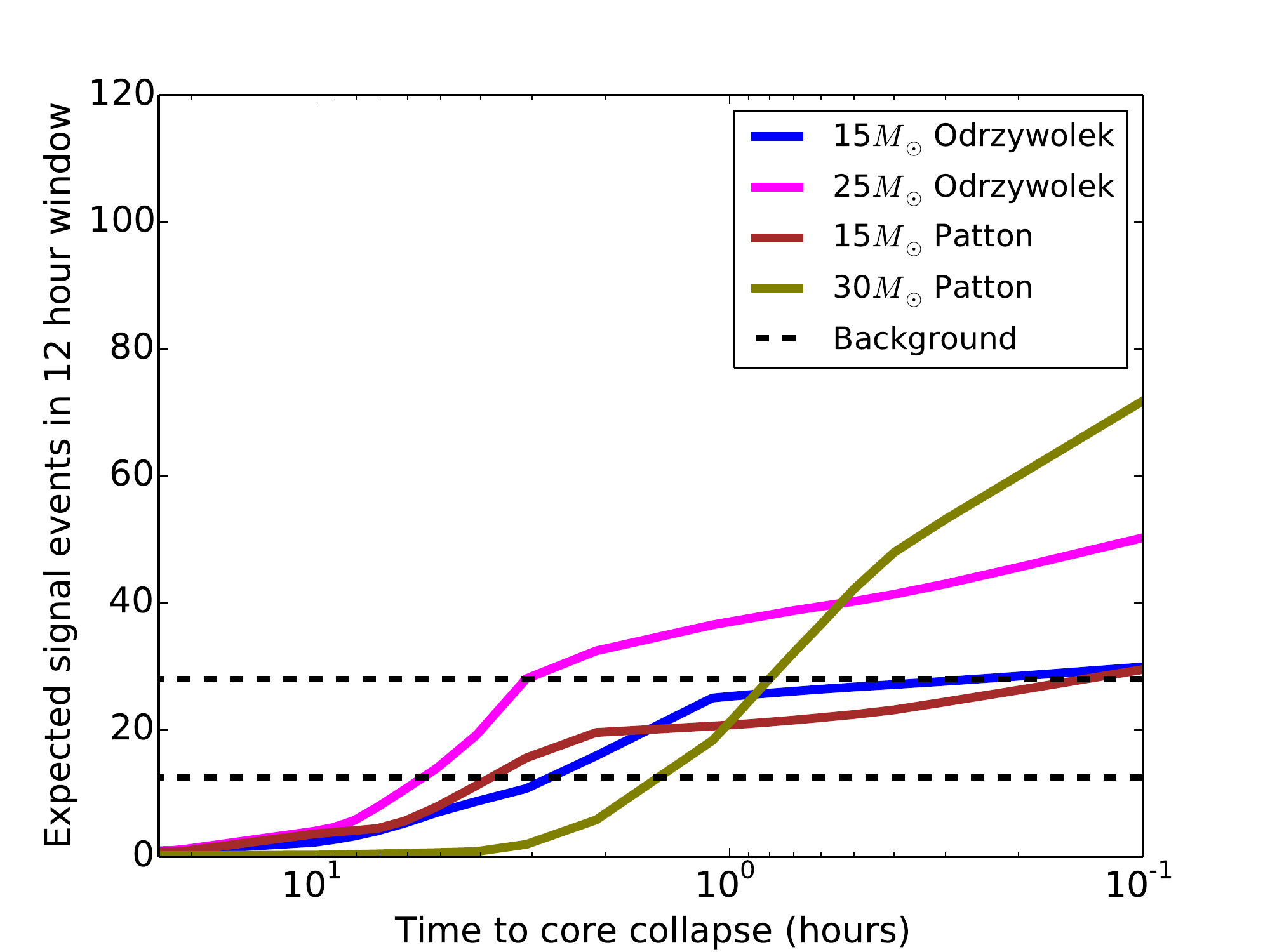}{0.49\textwidth}{(b) DC expected signal rate}
		}
		\caption{
			Expected signal events in 12~hour time window, after detection efficiencies are taken into account as per \autoref{Sec:Detec}. 
			A distance of 200~pc and NO is assumed.
			Dotted lines show the high and low background assumptions.
		}		\label{Fig:Rate_det}
	\end{figure}

	Discussions of pre-SN stars often focus on \alphaOri{} (Betelgeuse) as an example of a nearby massive star, although $\mathrm{\alpha}$-Sco(Antares) has a similar mass and distance. 
	For the purpose of this study \alphaOri{} is assumed to be 200~pc from Earth, with a mass between 15~and~25~\Msun{}.
	Estimates of Betelgeuse's mass are correlated with its distance (\cite{EvTrackBetelgeuse}), so two extremes chosen for benchmarking performance are that it is 150~pc away and 15~\Msun{}, or 250~pc away and 25~\Msun{}.  
	These values are chosen for comparison to \cite{Asakura:2015bga}, rather than to match the most up-to-date precise estimates of \alphaOri{}'s distance and mass.
	\autoref{Fig:Rate_det} shows the expected number of detected events at SK-Gd under these assumptions, after detection efficiencies are taken into account. 
	The model with 30~\Msun{} is also included for the sake of comparison.
	The expected number of detected events in the final 12~hours before collapse are summarised in \autoref{Tab:Ndetected}, and the expected amount of warning is summarised in \autoref{Tab:Betelgeuse_exp}. 
	\begin{table}
	\center
		\begin{tabular}{ccc|rcr|rcr} 
		&& Mass &  \multicolumn{3}{c}{Single}& \multicolumn{3}{c}{}\\
		&Model & (\Msun{})&\multicolumn{3}{c}{neutrons}& \multicolumn{3}{c}{DC}\\
		\hline
		NO 	& Odrzywolek & 15 & 55 & - & 71 	& 33	& - & 36 \\
			& Patton      & 15 & 65 & - & 84 	& 45	& - & 50  \\
			& Odrzywolek & 25 & 120 & - & 160 	& 59	& - & 65  \\
			& Patton      & 30 & 130 & - & 170 	& 100	& - & 110 \\
		
		\midrule
		IO 	& Odrzywolek & 15 & 16 & - & 20 & 9	& - & 10 \\
			& Patton      & 15 & 18 & - & 23 & 13	& - & 15 \\
			& Odrzywolek & 25 & 34 & - & 44 & 17	& - & 18 \\
			& Patton      & 30 & 40 & - & 52 & 34	& - & 37 \\
				\end{tabular}
		\caption{Expected numbers of events at 200~kpc in the final 12~hours before collapse at SK-Gd. Uncertainty comes from TNC \gammaray{} model only.}
		\label{Tab:Ndetected}
	\end{table}

	\begin{table}[htbp]
		\center
		\begin{tabular}{cccc|rcr|rcr}
		MO&&&Assumed&\multicolumn{6}{c}{Warning (hours)}\\
		&Mass&&distance&\multicolumn{6}{c}{given FPR}\\
		&(\Msun{})&Model&(pc)&\multicolumn{3}{c}{1/year} &\multicolumn{3}{c}{1/cy.} \\
		\toprule
		NO & 15 & 150& Odrzywolek & 5.3 &-& 8.4 & 3.4 &-& 6.3  \\
		 & 15 & 150& Patton & 7.1 &-& 14.1 & 5.1 &-& 9.8  \\
		 & 25 & 250& Odrzywolek & 4.7 &-& 7.4 & 3.3 &-& 5.7 \\
		 & 30 & 250& Patton & 1.0 &-& 1.6 & 0.7 &-& 1.1 \\
		\midrule
		IO & 15 & 150& Odrzywolek & 0.1 &-& 2.0 & 0.0 &-& 0.8 \\
		 & 15 & 150& Patton & 0.3 &-& 4.1 & 0.0 &-& 2.2 \\
		 & 25 & 250& Odrzywolek & 0.0 &-& 0.6 & 0.0 &-& 0.0\\
		 & 30 & 250& Patton & 0.1 &-& 0.4 & 0.0 &-& 0.1\\
			\bottomrule
		\end{tabular}
		\caption{Time at which expected signal exceeds threshold, for some assumptions chosen to represent Betelgeuse.
		30\Msun{} at 250~pc is far from the range of mass estimates for Betelgeuse, but is provided anyway for comparison.
		Uncertainty comes from TNC \gammaray{} model, and background uncertainty.}
		\label{Tab:Betelgeuse_exp}
	\end{table}

	\begin{figure}[htbp]
		\gridline{
			\fig{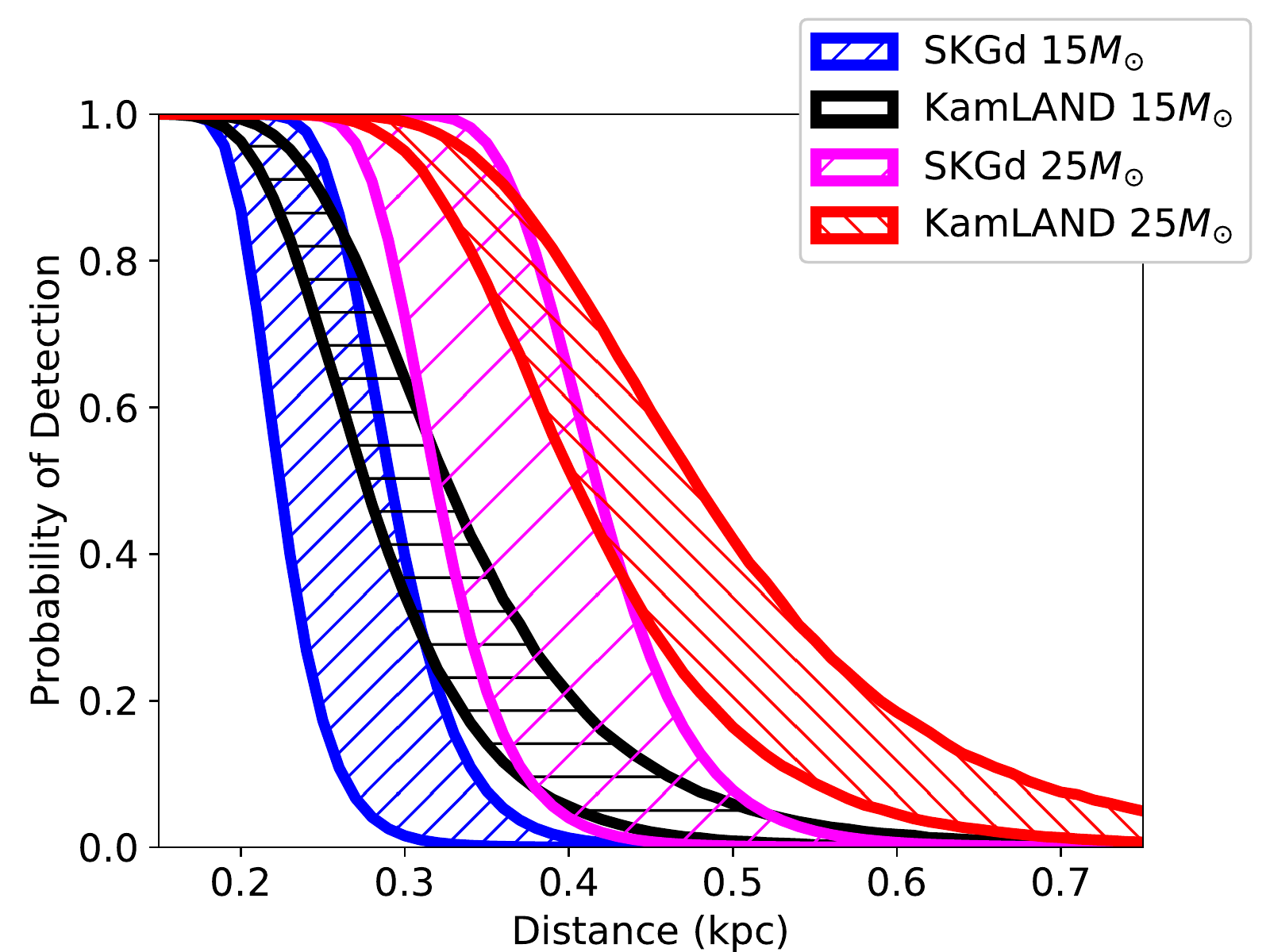}{0.49\textwidth}{(a) NO, 5~\tsigma{}/48 hours FPR equivalent}
			\fig{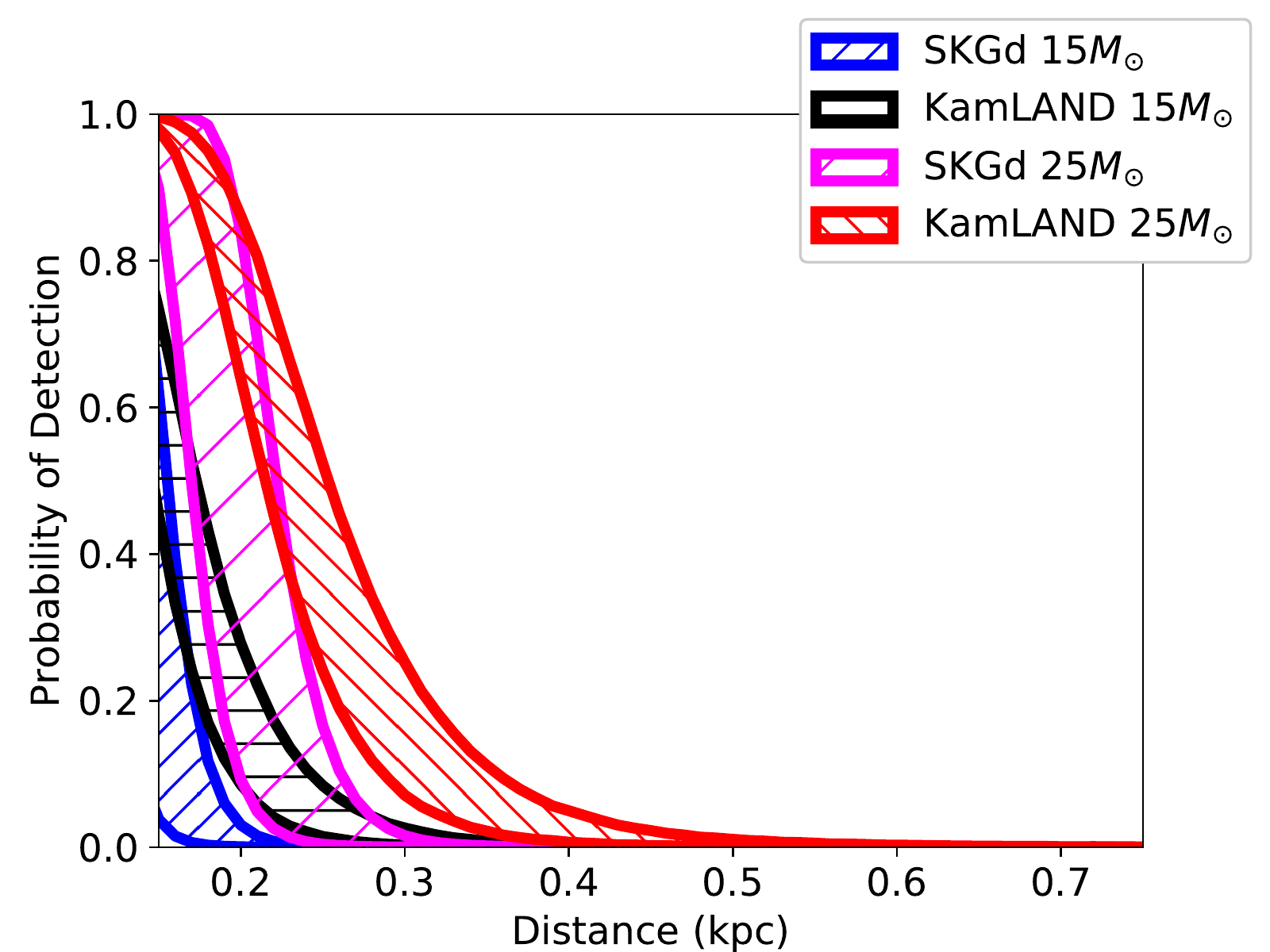}{0.49\textwidth}{(b) IO, 5~\tsigma{}/48 hours FPR equivalent}
		}
		\gridline{
			\fig{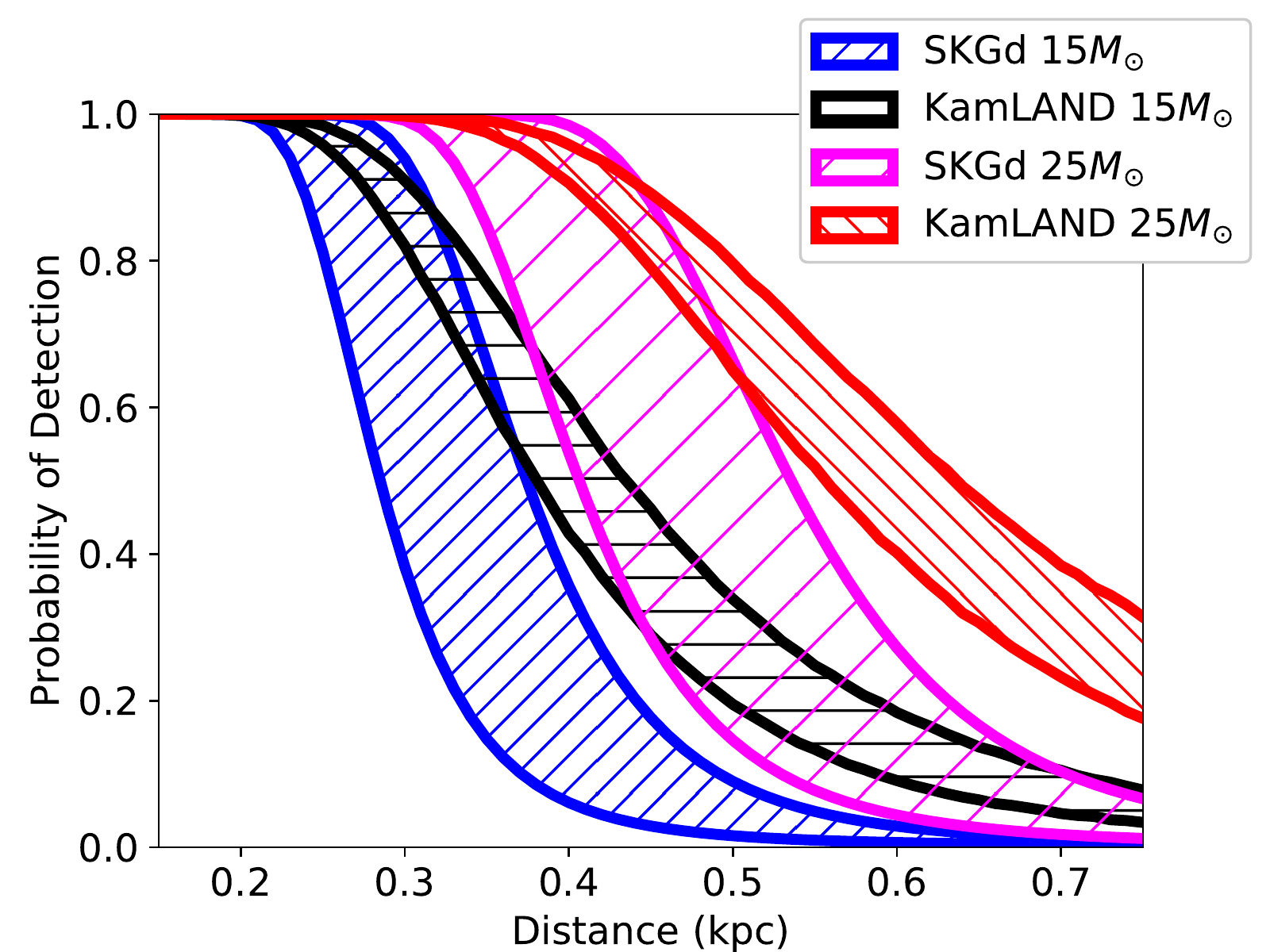}{0.49\textwidth}{(c) NO, 3~\tsigma{}/48 hours FPR equivalent}
			\fig{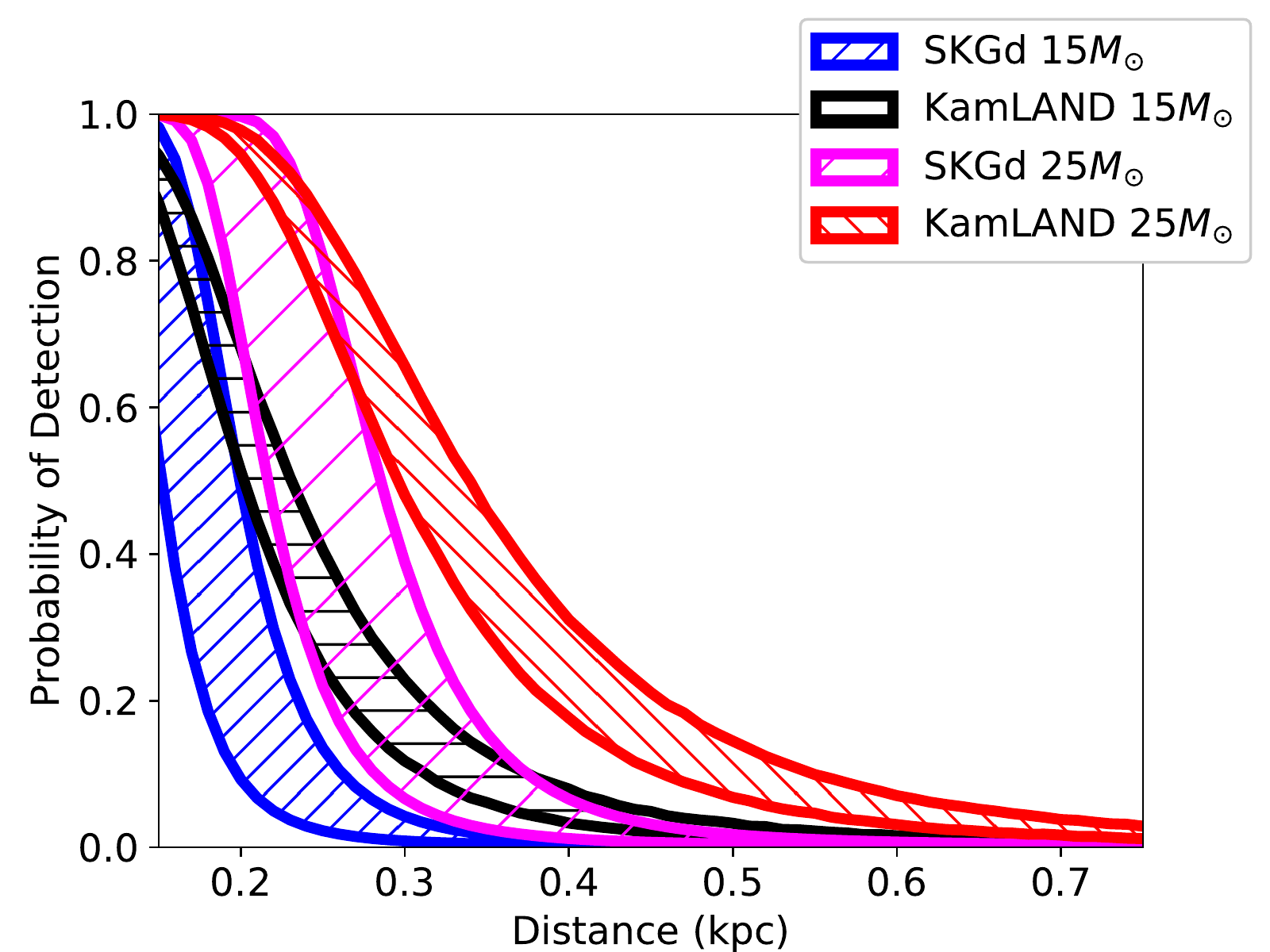}{0.49\textwidth}{(d) IO, 3~\tsigma{}/48 hours FPR equivalent}
		}		\caption{
		As \autoref{Fig:RangePatton}, but with reference numbers for KamLAND included.
		The FPR is fixed at the equivalent of of 5~\tsigma{}/3~\tsigma{} with a trial factor of 1~per~48~hours.
		Differences in expected alarm latency are not taken into account.
		Pre-SN models shown are those of Odrzywolek et al.
		}
		\label{Fig:Range_KL}
	\end{figure}
	The KamLAND collaboration published an analysis of their own sensitivity to pre-SN neutrinos (\cite{Asakura:2015bga}), which is compared to the expected sensitivity of SK-Gd. 
	KamLAND is a liquid scintillator detector based in the same mine as SK. 
	It has lower energy thresholds than SK, and so would detect IBD events from a pre-SN more efficiently, and has lower background rates. 
	However, the mass of SK is more than 20 times larger than that of KamLAND, so more events are seen in total.

	The nominal performance of KamLAND is taken from \cite{Asakura:2015bga}.
	KamLAND background rates are 0.071-0.355 events per day depending on Japanese nuclear reactor power. 
	Events are integrated over a 48~hour sliding window each 15~minutes. 
	Assumed signal rates in the final 48~hours at 200~pc are 25.7(7.28) in the 25~\Msun{} case, 12.0(3.38) in the 15~\Msun{} case, for the NO(IO) case.
	The pre-SN models used were those of \cite{OdrzywolekHeger}.
	Not enough information is provided in \cite{Asakura:2015bga} to directly and fairly compare warning times.
	
	\autoref{Fig:Range_KL} shows the probability of detection before core collapse (t=0) against distance to the pre-SN star.
	The estimated range for KamLAND is also shown. 
	KamLAND has a latency of 25~minutes, which is not taken into account.
	The FPR is set to match that of the 3~\tsigma{} and 5~\tsigma{} with a 48~hour signal window used by KamLAND, for the sake of comparison.
	That is, a the false positive rate is set to $\frac{1}{370}$~per~48~hours for $3$~$\sigma$ and $\frac{1}{1744278}$~per~48~hours for $5$~$\sigma$.
	By this comparison, the maximum detection range of SK-Gd is slightly shorter than that of KamLAND. 
	This is due to KamLAND's lower expected background rate.

	\added{
	Next generation liquid scintillator and Gd loaded water Cherenkov detectors could provide earlier warning to longer distances due to their large target masses.
	A future dark-matter direct-detection experiment could also detect significant numbers of pre-SN neutrinos through coherent scattering, with the advantage of being sensitive to all flavours (\cite{Raj:2019wpy}).}\explain{Reviewer comment:+ The paper compares SK-Gd's sensitivity with KamLAND's sensitivity. Please add some comments about preSN neutrino with direct dark matter detection experiments(﻿arXiv:1905.09283)}

\section{Conclusion} \label{Sec:Conclusion}
	Electron anti-neutrinos from a pre-SN star precede those from a CCSN by hours or days, increasing in flux and energy rapidly over a period of hours: 
	this has never been detected.
	In the next stage of SK, gadolinium loading will enable efficient identification of neutrons, enabling the reduction in the energy threshold for the detection of \tantinu{e}. 

	The background rates and signal efficiencies for an SK-Gd low energy analysis capable of detecting pre-SN \tantinu{e} have been quantified. 
	This requires detection of events below the usual energy thresholds of SK, for which trigger efficiency and reconstruction are poorer, and backgrounds higher.
	Gadolinium loading is essential to detecting these events.
	Through a rapid increase in the number of event candidates, additional warning of a very nearby SN can be achieved, and useful information provided about late stellar burning processes that lead up to a supernova.
	
	Based on this and the predicted fluxes of \cite{OdrzywolekHeger} and \cite{Patton:2017neq}, estimates were produced of the distance at which a pre-SN star could be observed, and the amount of additional early warning that could be expected. 
	Uncertainty in the future capabilities of the detector arises mainly from the future internal contamination of the SK detector, which is the main source of backgrounds at low energy.
	This uncertainty will be reduced once in-situ measurements of background become available.
	An inverted neutrino mass ordering would have a detrimental effect on the range of this technique by reducing the \tantinu{e} fraction of the \tantinu{} flux leaving the star. 

	The nearest red supergiant star to Earth is \alphaOri{}, which we assume to have an initial mass of 15-25\Msun{} and distance from Earth of 150-250~pc.
	Assuming normal neutrino mass ordering, 0.2\% gadolinium sulfate loading at SK, \alphaOri{} going pre-SN could lead to the detection of more than 200 events in SK-Gd in the final 12~hours before core collapse, well exceeding the expected background.
	Assuming a statistical false positive rate of 1~per~century, if it were pre-SN, \alphaOri{} could be detected 3 to 10~hours before core collapse, and the greatest distance at which a pre-SN star could be detected is 500~pc.
	Allowing a higher false positive rate of 1~per~year, 5 to 14~hours of early warning could be achieved for \alphaOri{}, and maximum detection range could extend to 600~pc.

	\replaced{Discussions have begun in the SK Collaboration as to how to implement a pre-SN alarm once Gd is loaded.}{A pre-SN alert could be provided by SK to the astrophysics community following gadolinium loading.}\explain{Reviewer comment: + This paper focus on the sensitivity. If authors add the clear message about that SK-Gd will make alert system for the the astrophysical community, it is better.}
	Future large neutrino detectors will improve the potential range of detection, especially if they have a sufficiently small low energy threshold and the ability to tag neutrons from IBD.
	It could also be possible to use multiple detectors in combination to provide a pre-SN alert with higher confidence.

\acknowledgments{
\section*{Acknowledgments}
	We gratefully acknowledge the cooperation of the Kamioka Mining and Smelting Company.
	The Super‐Kamiokande experiment has been built and operated from funding by the 
	Japanese Ministry of Education, Culture, Sports, Science and Technology, the U.S.
	Department of Energy, and the U.S. National Science Foundation. Some of us have been 
	supported by funds from the National Research Foundation of Korea NRF‐2009‐0083526
	(KNRC) funded by the Ministry of Science, ICT, and Future Planning and the Ministry of
	Education (2018R1D1A3B07050696, 2018R1D1A1B07049158), 
	the Japan Society for the Promotion of Science, the National
	Natural Science Foundation of China under Grants No. 11235006, the National Science and 
	Engineering Research Council (NSERC) of Canada, the Scinet and Westgrid consortia of
	Compute Canada, the National Science Centre, Poland (2015/18/E/ST2/00758),
	the Science and Technology Facilities Council (STFC) and GridPP, UK, 
	and the European Union’s H2020-MSCA-RISE-2018 JENNIFER2 grant agreement no.822070.
	}

\clearpage
\bibliographystyle{aastex.bst}
\bibliography{main.bib}

\listofchanges
\end{document}